\documentclass[a4paper,twocolumn,floatfix,superscriptaddress,longbibliography,accepted=2023-11-15]{quantumarticle}

\pdfoutput=1
\usepackage{amssymb,amsmath,amstext}
\usepackage{graphicx}
\usepackage{epstopdf}
\usepackage{color}  
\usepackage{xcolor}
\usepackage{appendix}
\usepackage[T1]{fontenc}
\usepackage{bbold}
\usepackage{bbm}
\usepackage{latexsym}
\usepackage[numbers,sort&compress]{natbib}
\usepackage[latin1]{inputenc}
\usepackage{tikz}
\usepackage{xspace}
\usepackage{amsthm}
\usepackage{standalone}
\usepackage{xr-hyper}

\usepackage{afterpage}
\usetikzlibrary{shapes,arrows}
\renewcommand{\vec}[1]{\boldsymbol{#1}}  

\long\def\ca#1\cb{} 

\newcommand{\braket}[2]{\langle #1 \hspace{1pt} | \hspace{1pt} #2 \rangle}

\newcommand{\norm}[2][]{#1| \! #1| #2 #1| \! #1|}

\newcommand{\ket}[1]{|#1\rangle}               
\newcommand{\bra}[1]{\langle #1|}              
\newcommand{\dya}[1]{\ket{#1}\!\bra{#1}}
\newcommand{\ip}[2]{\langle #1|#2\rangle}      
\newcommand{\matl}[3]{\langle #1|#2|#3\rangle} 

\newcommand{\SC}{\mathcal{S}}

\newcommand{\Tr}{{\rm Tr}}

\newcommand{\ave}[1]{\langle #1\rangle}               
\renewcommand{\geq}{\geqslant}
\renewcommand{\leq}{\leqslant}
\newcommand{\mte}[2]{\langle#1|#2|#1\rangle }

\renewcommand{\Re}{\text{Re}}
\renewcommand{\Im}{\text{Im}}

\newcommand{\CU}{\widehat{C}}
\newcommand{\CN}{C}

\newcommand{\opt}{\text{opt}}

\DeclareMathOperator*{\argmax}{arg\,max}
\DeclareMathOperator*{\argmin}{arg\,min}
\renewcommand{\vec}[1]{\boldsymbol{#1}}  
\newcommand{\ot}{\otimes}
\newcommand{\ad}{^\dagger}

\newcommand*{\id}{\openone}

\newtheorem{proposition}{Proposition}
\newtheorem{proposition2}{Proposition}
\newcommand{\DQC}{\ensuremath{\mathsf{DQC1}}\xspace}
\newcommand{\poly}{\operatorname{poly}}

\begin{document}

\usetikzlibrary{calc}

\tikzstyle{decision} = [diamond, draw, fill=purple!60, 
    text width=4.5em, text badly centered, node distance=3cm, inner sep=0pt, rounded corners, minimum height=7em, minimum width = 7em]
\tikzstyle{block} = [rectangle, draw, fill=blue!60, 
    text width=5em, text centered, rounded corners, minimum height=4em]
\tikzstyle{block2} = [rectangle, draw, fill=orange!60, 
    text width=5em, text centered, rounded corners, minimum height=4em]
\tikzstyle{line} = [draw, -latex']
\tikzstyle{cloud} = [draw, ellipse, fill=pink!150, node distance=3cm,
    minimum height=2.5em]

\title{Variational Quantum Linear Solver}

\author{Carlos Bravo-Prieto}
\affiliation{Theoretical Division, Los Alamos National Laboratory, Los Alamos, NM 87545, USA.}
\affiliation{Barcelona Supercomputing Center, Barcelona, Spain.}
\affiliation{Institut de Ci\`encies del Cosmos, Universitat de Barcelona, Barcelona, Spain.}

\author{Ryan LaRose}
\affiliation{Department of Computational Mathematics, Science, and Engineering \& Department of Physics and Astronomy, Michigan State University, East Lansing, MI 48823, USA.}

\author{M. Cerezo} 
\affiliation{Theoretical Division, Los Alamos National Laboratory, Los Alamos, NM 87545, USA.}
\affiliation{Center for Nonlinear Studies, Los Alamos National Laboratory, Los Alamos, NM, USA
}

\author{Yi\u{g}it Suba\c{s}\i}
\affiliation{Computer, Computational and Statistical Sciences Division, Los Alamos National Laboratory, Los Alamos, NM 87545, USA}

\author{Lukasz Cincio} 
\affiliation{Theoretical Division, Los Alamos National Laboratory, Los Alamos, NM 87545, USA.}

\author{Patrick J. Coles}
\affiliation{Theoretical Division, Los Alamos National Laboratory, Los Alamos, NM 87545, USA.}

\begin{abstract}
Previously proposed quantum algorithms for solving linear systems of equations cannot be implemented in the near term due to the required circuit depth. Here, we propose a hybrid quantum-classical algorithm, called Variational Quantum Linear Solver (VQLS), for solving linear systems on near-term quantum computers. VQLS seeks to variationally prepare $|x\rangle$ such that $A|x\rangle\propto|b\rangle$.
We derive an operationally meaningful termination condition for VQLS that allows one to guarantee that a desired solution precision $\epsilon$ is achieved. Specifically, we prove that $C \geq \epsilon^2 / \kappa^2$, where $C$ is the VQLS cost function and $\kappa$ is the condition number of $A$. We present efficient quantum circuits to estimate $C$, while providing evidence for the classical hardness of its estimation. Using Rigetti's quantum computer, we successfully implement VQLS up to a problem size of $1024\times1024$. Finally, we numerically solve non-trivial problems of size up to $2^{50}\times2^{50}$. For the specific examples that we consider, we heuristically find that the time complexity of VQLS scales efficiently in $\epsilon$, $\kappa$, and the system size $N$.
\end{abstract}

\maketitle

\section{Introduction} 

Linear systems of equations play an important role in many areas of science and technology, including machine learning~\cite{alpaydin2010, bishop2006},
solving partial differential equations~\cite{evans2010},
fitting polynomial curves~\cite{bretscher1995}, and analyzing electrical circuits~\cite{spielman2011}.
In the past decade, significant attention has been given to the possibility of solving linear systems on quantum computers. Classically solving an $N \times N$ linear system ($N$ equations for $N$ unknowns) scales polynomially in $N$. In contrast, Harrow-Hassidim-Lloyd (HHL) introduced a quantum algorithm that scales logarithmically in $N$, suggesting that quantum computers may provide an exponential speedup for certain linear system problems~\cite{HHL}. More precisely, the HHL algorithm treats the Quantum Linear Systems Problem (QLSP), where the goal is to prepare a quantum state $\ket{x}$ that is proportional to a vector $\vec{x}$ that satisfies the equation $A\vec{x} = \vec{b}$. If both $A$ and $\vec{b}$ are sparse, then for a fixed precision $\epsilon$ in the solution, the complexity of HHL scales polynomially in $\log N$ and $\kappa$, where $\kappa$ is the condition number of $A$, i.e, the ratio of the largest  to the smallest singular value. Further improvements to HHL have reduced the complexity to linear $\kappa$ scaling~\cite{Ambainis,Yigit} and polylogarithmic scaling in $1/\epsilon$~\cite{CKS,Chakraborty}, as well as improved the sparsity requirements~\cite{Wossnig}.

The aforementioned quantum algorithms hold promise for the future,  when large-scale quantum computers exist with enough qubits for quantum error correction. The timescale for such computers remains an open question, but is typically estimated to be on the order of two decades. On the other hand, commercial quantum computers currently exist with $\sim 50$ noisy qubits, with the number of qubits rapidly increasing. A crucial question is how to make use of such noisy intermediate-scale quantum (NISQ) computers~\cite{preskill2018quantum}. In principle, one can implement the aforementioned quantum algorithms on NISQ devices, however noise limits the problem size to be extremely small. For example, the HHL algorithm has been implemented with superconducting qubits~\cite{Zheng,lee2019hybrid}, nuclear magnetic resonance (NMR)~\cite{Pan}, and photonic devices~\cite{Cai, Barz}, but these experiments were limited to a problem size of $2 \times 2$. More recently, an alternative approach based on an adiabatic-inspired quantum algorithm~\cite{Yigit} was implemented with NMR for an $8 \times 8$ problem, and this appears to be the current record for the largest linear system solved with a gate-based quantum computer~\cite{Wen}.

\begin{figure*}[t]
	\centering
	\includegraphics[scale=0.90]{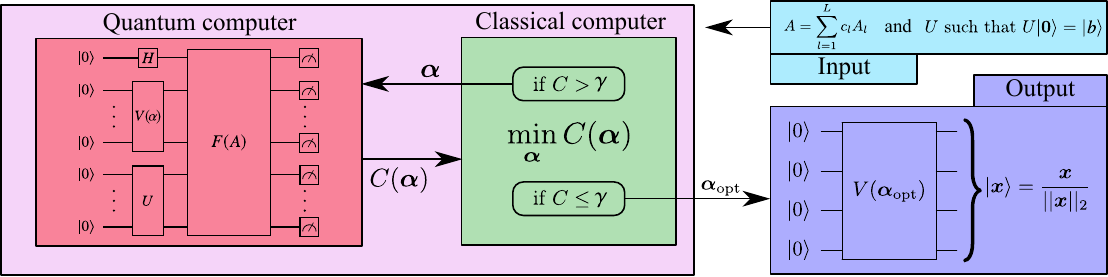}
	\caption{Schematic diagram for the VQLS algorithm. The input to VQLS is a matrix $A$ written as a linear combination of unitaries $A_l$ and a short-depth quantum circuit $U$ which prepares the state $\ket{b}$. The output of VQLS is a quantum state $\ket{x}$ that is approximately proportional to the solution of the linear system $A \vec{x} = \vec{b}$. Parameters $\vec{\alpha}$ in the ansatz $V(\vec{\alpha})$ are adjusted in a hybrid quantum-classical optimization loop until the cost $C(\vec{\alpha})$ (local or global) is below a user-specified threshold. When this loop terminates, the resulting gate sequence $V(\vec{\alpha}_\text{opt})$ prepares the state $\ket{x} = \vec{x} / ||\vec{x}||_2$, from which observable quantities can be computed. Furthermore, the final value of the cost $C(\vec{\alpha}_\text{opt})$ provides an upper bound on the deviation of observables measured on $\ket{x}$ from observables measured on the exact solution.} 
	\label{fig:VQLS_Flowchart}
\end{figure*}

An interesting strategy to make use of NISQ devices is to employ variational hybrid quantum-classical algorithms (VHQCAs). VHQCAs manage to reduce quantum circuit depth at the expense of additional classical optimization. Specifically, VHQCAs employ a short-depth quantum circuit to efficiently evaluate a cost function, which depends on the parameters of a quantum gate sequence, and then leverage well-established classical optimizers to minimize this cost function. For example, while Shor's algorithm for factoring is not a near-term algorithm, recently a VHQCA for factoring was introduced potentially making factoring nearer term~\cite{VQF}. Other VHQCAs have been proposed for chemistry~\cite{VQE,cao2018quantum, VQEex, Jones}, simulation~\cite{Li, Kokail,heya2019subspace,cirstoiu2019variational,yuan2019theory}, data compression~\cite{Romero}, state diagonalization~\cite{VQSD, VQSVD,cerezo2020variational}, compiling~\cite{QAQC,jones2018quantum}, quantum foundations~\cite{arrasmith2019variational}, fidelity estimation~\cite{cerezo2019variational}, and metrology~\cite{koczor2019variational}.

In this work, we propose a VHQCA for solving the QLSP. Our algorithm, called the Variational Quantum Linear Solver (VQLS), employs a cost function that quantifies either the global or local closeness of the quantum states $A\ket{x} $ and $\ket{b}$, where the latter is a normalized version of $\vec{b}$. We provide efficient quantum circuits to estimate our cost functions, and show under typical complexity assumptions that they cannot be efficiently estimated classically. Furthermore, we derive operational meaning for our cost functions, as upper bounds on $\epsilon^2 / \kappa^2$. This is crucial since it gives a termination criterion for VQLS that guarantees a desired precision $\epsilon$.

It is important to emphasize that all VHQCAs are heuristic algorithms, making rigorous complexity analysis  difficult. Nevertheless, our numerical simulations (without finite sampling, and both for specific $A$ and for randomly chosen $A$) indicate that the run time of VQLS scales efficiently in $\kappa$, $\epsilon$, and $N$. Namely, we find evidence of (at worst) linear scaling in $\kappa$, logarithmic scaling in $1/\epsilon$, and polylogarithmic scaling in $N$.

We employ Rigetti's Quantum Cloud Services to implement VQLS. With their quantum hardware, we were able to successfully solve a particular linear system of size $1024 \times 1024$. We are therefore optimistic that VQLS could provide a near-term approach to the QLSP.

\section{Results} 

\subsection{VQLS Algorithm}

\subsubsection{Overview}

Figure~\ref{fig:VQLS_Flowchart} shows a schematic diagram of the VQLS algorithm. 
The input to VQLS is: (1) an efficient gate sequence $U$ that prepares a quantum state $\ket{b}$ that is proportional to the vector $\vec{b}$, and (2) a decomposition of the matrix $A$ into a linear combination of $L$ unitaries of the form 
\begin{equation}\label{eq-A-decom}
A = \sum_{l=1}^{L} c_l A_l\,,
\end{equation}
where the $A_l$ are unitaries, and the $c_l$ are complex numbers.  The assumption that $A$ is given in this form is analogous to the assumption that the Hamiltonian $H$ in the variational quantum eigensolver~\cite{VQE} is given as a linear combination of Pauli operators $H=\sum_{l=1}^{L} c_l \sigma_l$, where naturally one makes the assumption that $L$ is only a polynomial function of the number of qubits, $n$. Additionally, we assume $\kappa<\infty$ and $\norm{A}\leq 1$, and that the $A_l$ unitaries can be implemented with efficient quantum circuits. Appendix~\ref{sec:sparse} describes an efficient method to decompose $A$ in the form of~\eqref{eq-A-decom} for the case when $A$ is a sparse matrix.

With this input, the Quantum Linear Systems Problem (QLSP) is to prepare a state $\ket{x}$ such that $A \ket{x}$ is proportional to $\ket{b}$. To solve this problem, VQLS employs an ansatz for the gate sequence $V(\vec{\alpha})$ that prepares a potential solution $\ket{x(\vec{\alpha})} =V(\vec{\alpha})\ket{\vec{0}} $. The parameters $\vec{\alpha}$ are input to a quantum computer, which prepares $\ket{x(\vec{\alpha})}$ and runs an efficient quantum circuit that estimates a cost function $C(\vec{\alpha})$. The precise details of the cost function and its estimation are discussed below. We simply remark here that $C(\vec{\alpha})$ quantifies how much component $A \ket{x}$ has orthogonal to $\ket{b}$. The value of $C(\vec{\alpha})$ from the quantum computer is returned to the classical computer which then adjusts $\vec{\alpha}$ (via a classical optimization algorithm) in an attempt to reduce the cost. This process is iterated many times until one reaches a termination condition of the form $C(\vec{\alpha})\leq \gamma$, at which point we say that $\vec{\alpha} = \vec{\alpha}_{\opt}$. 

VQLS outputs the parameters $\vec{\alpha}_{\opt}$, which can then be used to prepare the quantum state $\ket{x(\vec{\alpha}_{\opt})} = V(\vec{\alpha}_{\opt}) \ket{\vec{0}}$. One can then measure observables of interest on the state $\ket{x(\vec{\alpha}_{\opt})}$ in order to characterize the solution vector. Due to the operational meaning of our cost function (discussed below), one can upper bound the deviation of observable expectation values for $\ket{x(\vec{\alpha}_{\opt})}$ from those of the true solution, based on the value of the cost function. Hence, before running VQLS, one can decide on a desired error tolerance $\epsilon$, where 
\begin{equation}\label{eq-epsilon}
 \epsilon = (1/2)\Tr |\hspace{4pt} \dya{x_0} - \dya{x(\vec{\alpha}_{\opt})}  \hspace{4pt} |
\end{equation}
is the trace distance between the exact solution $\ket{x_0}$ and the approximate solution $\ket{x(\vec{\alpha}_{\opt})}$. This $\epsilon$ then translates into a threshold value $\gamma$ that the final cost $C(\vec{\alpha}_{\opt})$ must achieve (see \eqref{eq-costfunctionsbounds}  for the relation between $\epsilon$ and $\gamma$).

\subsubsection{Cost functions}

For simplicity, we write $\ket{x(\vec{\alpha})}$ as $\ket{x}$ henceforth. Here we discuss several reasonable cost functions. A simple, intuitive cost function involves the overlap between the (unnormalized) projector $\dya{\psi}$, with $\ket{\psi} = A\ket{x}$, and the subspace orthogonal to $\ket{b}$, as follows: 
\begin{equation} \label{eqn:global-cost1}
 \CU_G = \Tr(\dya{\psi}(\id - \dya{b})) = \mte{x}{H_G}\,.    
\end{equation}
We note that one can view this cost function as the expectation value of an effective Hamiltonian~ 
\begin{equation} \label{eqn:hamiltonian-of-global-cost}
 H_G = A\ad (\id - \dya{b}) A\,,
\end{equation}
which is similar to the final Hamiltonian in Ref.~\cite{Yigit}. The $\CU_G$ function is small if $\ket{\psi}$ is nearly proportional to $\ket{b}$ or if the norm of $\ket{\psi}$ is small. The latter does not represent a true solution, and hence to deal with this, one can divide $\CU_G$ by the norm of $\ket{\psi}$ to obtain
\begin{equation} \label{eqn:global-cost2}
 \CN_G = \CU_G / \ip{\psi}{\psi} = 1 - |\ip{b}{\Psi}|^2\,,  
\end{equation}
where $\ket{\Psi} = \ket{\psi} / \sqrt{\ip{\psi}{\psi}}$ is a normalized state. As discussed in the Supplemental Material, $\CN_G$ and $\CU_G$ have similar performance for the QLSPs that we considered.

We emphasize that global cost functions such as those in \eqref{eqn:global-cost1} and \eqref{eqn:global-cost2} can exhibit barren plateaus, i.e., cost function  gradients that vanish exponentially in the number of qubits $n$, see Ref.~\cite{cerezo2020cost}. To improve trainability for large $n$, one can introduce local versions of these costs, as follows:
\begin{align} \label{eqn:local-cost1}
 \CU_L =  \mte{x}{H_L}\,,\qquad  \CN_L = \CU_L / \ip{\psi}{\psi}\,, 
\end{align}
where the effective Hamiltonian is
\begin{equation}\label{eqn:hamiltonian-of-local-cost}
 H_L = A\ad U \left( \id - \frac{1}{n}\sum_{j=1}^n \dya{0_j} \otimes \id_{\overline{j}} \right) U\ad A \,,
 \end{equation}
with $\ket{0_j}$ the zero state on qubit $j$ and $\id_{\overline{j}}$ the identity on all qubits except qubit $j$. One can show that (see Appendix~\ref{sec:faith})
\begin{align}
\label{eqnCostEquivalence}
\CU_L \leq \CU_G \leq n \CU_L\,,\qquad
\CN_L \leq \CN_G \leq n \CN_L\,,
\end{align}
which implies that $\CU_L = 0 \leftrightarrow \CU_G = 0$ and $\CN_L = 0 \leftrightarrow \CN_G = 0$. We assume that $\kappa$ is not infinite (i.e., that $A$ is full rank) and hence that $\ip{\psi}{\psi} \neq 0$. This implies that all four cost functions vanish under precisely the same conditions, namely, when $\ket{\psi} \propto \ket{b}$, which is the case when $\ket{x}$ is a solution to the QLSP.

\begin{figure}[t]
	\centering
	\includegraphics[width= \columnwidth]{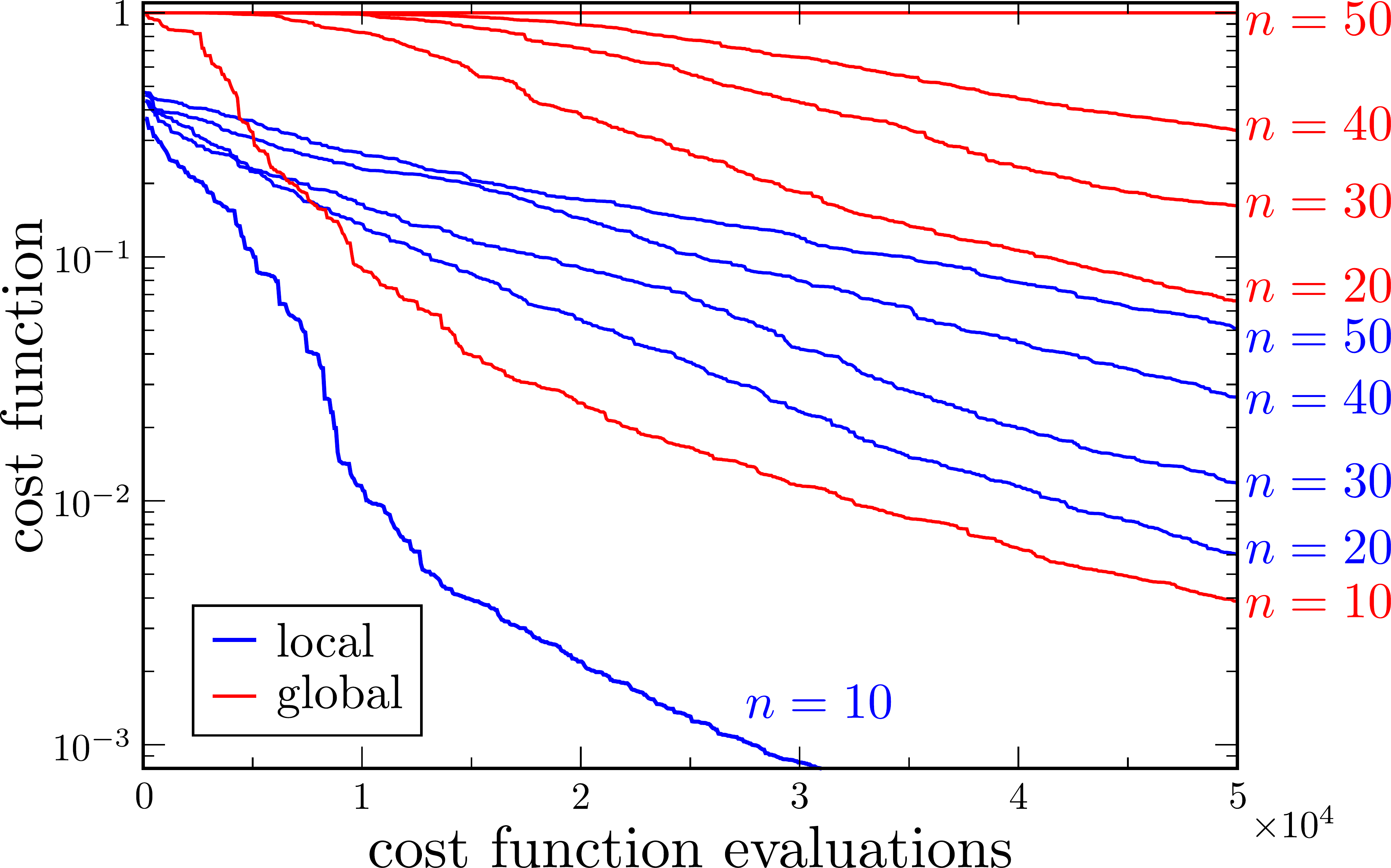}
	\caption{Comparison of local $\CN_L$ and global $\CN_G$ cost performance. Here we consider the QLSP of Eq.~\eqref{eqn:ExampleQLSP} for different system sizes. In all cases $\kappa=20$. For each $n \in \{10,\ldots,50\}$, we plot the cost value versus the number of cost function evaluations. As $n$ increases it becomes increasingly hard to train to global cost function. At $n=50$, our optimization cannot significantly lower $\CN_G$ below a value of one. On the other hand, we are able to train  $\CN_L$ for all values of $n$ considered.  \label{fig:local_vs_global}}
\end{figure}

As shown in Fig.~\ref{fig:local_vs_global}, as $n$ increases it becomes increasingly hard to optimize the global cost function $\CN_G$. On the other hand, the local cost function $\CN_L$ performs significantly better, as we are able to train $\CN_L$ for systems of size up to $2^{50}\times 2^{50}$ (i.e., with $50$ qubits). These results show that the vanishing gradients of global cost functions could make them untrainable for large $n$, and hence we propose using our local cost functions for large-scale implementations.

\subsubsection{Operational meaning of cost functions}\label{sec:OpMeaning}

Here we provide operational meanings for the aforementioned cost functions. These operational meanings are crucial since they allow one to define termination conditions for VQLS in order to achieve a desired precision. In particular, we find that the following bounds hold in general:
\begin{equation} \label{eq-costfunctionsbounds}
  \CU_G \geq \frac{\epsilon^2}{\kappa^2} \,, \,\,\, \CN_G \geq \frac{\epsilon^2}{\kappa^2} \,, \,\,\,  \CU_L \geq \frac{1}{n} \frac{\epsilon^2}{\kappa^2} \,, \,\,\, \CN_L \geq \frac{1}{n} \frac{\epsilon^2}{\kappa^2}\,.
\end{equation}
Note that one can take the right-hand-sides of these inequalities as the $\gamma$ quantity shown in Fig.~\ref{fig:VQLS_Flowchart}. 

We remark that, for $\CN_G$ and $\CN_L$, the bounds in \eqref{eq-costfunctionsbounds} can be tightened (by using the bounds on $\CU_G$ and $\CU_L$ in \eqref{eq-costfunctionsbounds}) as follows:
\begin{equation}
\label{eq-costfunctionsboundsTightened}
    \CN_G \geq \frac{\epsilon^2}{\kappa^2 \braket{\psi}{\psi}} \ , \quad
    \CN_L \geq \frac{1}{n} \frac{\epsilon^2}{\kappa^2 \braket{\psi}{\psi}} \ .
\end{equation}
Here, $\braket{\psi}{\psi}$ is experimentally computable (see \eqref{eq-psipsi} below) and satisfies $\braket{\psi}{\psi}\leq 1$. Hence, when training $\CN_G$ or $\CN_L$, one can employ the right-hand-sides of \eqref{eq-costfunctionsboundsTightened} as opposed to those of \eqref{eq-costfunctionsbounds} as the termination condition $\gamma$.

Furthermore, one can employ the operational meaning of the trace distance~\cite{nielsen_chuang} to note that, for any POVM element $M$,   we have $\epsilon \geq D(M)$, where
\begin{equation}
\label{eqn:DMdefinition}
D(M) = | \mte{x}{M} - \mte{x_0}{M} |   
\end{equation}
measures the difference between expectation values on $\ket{x}$ and $\ket{x_0}$. Relaxing to the general case where $M$ is any Hermitian observable gives $\epsilon \geq D(M)/(2\|M\|)$, and hence \eqref{eq-costfunctionsbounds} is a bound on observable differences.

Let us now provide a proof for \eqref{eq-costfunctionsbounds}. Consider first that $\CU_G = \ave{H_G}$, with the eigenstates and eigenvalues of $H_G$ denoted by $\{\ket{x_i}\}$ and $\{E_i\}$, respectively for $i=0,1,\dots$. By construction $\ket{x_0}$ is the ground state of $H_G$ with $E_0=~0$. In what follows we assume for simplicity that $\ket{x_1}$ is non-degenerate, although the same proof approach works for the degenerate case. 

It is clear that for a given $\epsilon$, the smallest energy $\ave{H_G}$ (hence cost) is achieved if the state $\ket{x}$ is a superposition of $\ket{x_0}$ and $\ket{x_1}$ only. One can see this by expanding an arbitrary state $\ket{x}$ in the energy eigenbasis, $\ket{x} = \sum_i \chi_i \ket{x_i}$, and noting that $\epsilon$ depends only on the magnitude of $\chi_0$. Hence for a fixed $\epsilon$, one is free to vary the set of coefficients $\{\chi_i\}_{i\neq 0}$, and the set that minimizes the energy corresponds to choosing $\chi_i = 0$ for all $i > 1$.

So we take:
\begin{align}
    \ket{x} = \cos(\theta/2) \ket{x_0} + e^{i\phi}\sin(\theta/2) \ket{x_1}\,,
\end{align}
and the associated energy is given by
\begin{align}
\label{eq:HG1}
    \mte{x}{H_G} =  E_1 \sin^2(\theta/2)  \geq \frac{\sin^2(\theta/2)}{\kappa^2}\,,
\end{align}
where we used the fact that $E_0=0$, and that the first excited state energy satisfies $E_1\geq 1/\kappa^2$ (which was shown in Ref.~\cite{Yigit}).
The trace distance between  $\ket{x}$ and $\ket{x_0}$ can be  easily computed as $\sqrt{1-|\ip{x}{x_0}|^2}$, which results in $ \epsilon=\vert \sin(\theta/2) \vert$. Inserting this into~\eqref{eq:HG1} yields $\CU_G \geq \epsilon^2/ \kappa^2$. The remaining inequalities in \eqref{eq-costfunctionsbounds} follow from \eqref{eqnCostEquivalence} and from the fact that $\ip{\psi}{\psi}\leq 1$, which implies $\CN_G\geq\CU_G$.

\subsubsection{Cost evaluation}

In principle, all the aforementioned cost functions can be efficiently evaluated using the Hadamard Test circuit and simple classical post-processing. However, in practice, care must be taken to minimize the number of controlled operations in these circuits. Consider evaluating the term $\ip{\psi}{\psi}$, which can be written as
\begin{align}\label{eq-psipsi}
    \ip{\psi}{\psi} = \sum_{ll'} c_l c_{l'}^* \beta_{ll'}\,,
\end{align}    
with
\begin{align}\label{eq-betall}
\beta_{ll'}=\mte{\vec{0}}{V\ad A_{l'}\ad A_{l} V }. 
\end{align}
There are $L(L-1)/2$ different $\beta_{ll'}$ terms that one needs to estimate, and which can be measured with Hadamard Tests. The Hadamard Test involves acting with $V$ on $\ket{\vec{0}}$, and then using an ancilla as the control qubit, applying $C_{A_l}$ followed by $C_{A_{l'}\ad}$, where $C_W$ denotes controlled-$W$ (see Appendix for precise circuits). 

In addition, for $\CU_G$ and $\CN_G$, one needs to evaluate 
\begin{align}
    |\ip{b}{\psi}|^2 = |\mte{\vec{0}}{U\ad A V}|^2 = \sum_{ll'} c_l c_{l'}^* \gamma_{ll'}\,,
\end{align}  
with
 \begin{align}   \label{eq-gamall}
    \gamma_{ll'} =\mte{\vec{0}}{U\ad A_l V} \mte{\vec{0}}{V\ad A_{l'}\ad U} . 
\end{align}
The $\gamma_{ll}$ terms are easily estimated by applying $U\ad A_l V$ to $\ket{\vec{0}}$ and then measuring the probability of the all-zeros outcome. For the $L(L-1)/2$ terms with $l\neq l'$, there are various strategies to estimate $\gamma_{ll'}$. For example, one could estimate the $L$ terms of the form $\mte{\vec{0}}{U\ad A_l V}$ with a Hadamard Test, but one would have to control all of the unitaries: $V$, $A_l$, and $U\ad$. Instead, we introduce a novel circuit called the Hadamard-Overlap Test that directly computes $\gamma_{ll'}$ without having to control $V$ or $U$ at the expense of doubling the number of qubits. This circuit is schematically shown in Fig.~\ref{fig:VQLS_Flowchart} and explained in detail in  Appendix~\ref{Sec:cost-circuite}.

Finally, for $\CU_L$ and $\CN_L$, one needs to estimate terms of the form
\begin{equation} \label{eq-deltall}
    \delta_{ll'}^{(j)}= \mte{\vec{0}}{V\ad A_{l'}\ad U (\dya{0_j} \otimes \id_{\overline{j}} ) U\ad A_l V}\,.
\end{equation}
These terms can either be estimated with the Hadamard-Overlap Test or with the Hadamard Test, which are discussed in  Appendix~\ref{Sec:cost-circuite}.

\subsubsection{Classical hardness of computing the cost functions}

Here we state that computing the cost functions in \eqref{eqn:global-cost1}, \eqref{eqn:global-cost2}, and \eqref{eqn:local-cost1} is classically hard under typical complexity assumptions. As shown in Appendix~\ref{App-DQC1}, the following proposition holds:
\begin{proposition}\label{Prop1}
The problem of estimating the VQLS cost functions  $\CU_G$, $\CN_G$, $\CU_L$, or $\CN_L$ to within precision $\pm \delta = 1/\poly(n)$ is \DQC-hard.
\end{proposition}
Recall that the complexity class Deterministic Quantum Computing with 1 Clean Qubit (\DQC) consists of all problems that can be efficiently solved with bounded error in the one-clean-qubit model of computation~\cite{knill1998POOQ}. Moreover, classically simulating \DQC~is impossible unless the polynomial hierarchy collapses to the second level~\cite{DBLP:journals/corr/FujiiKMNTT14, 1704.03640}, which is not believed to be the case. Hence, Proposition \ref{Prop1} strongly suggests that a classical algorithm cannot efficiently estimate the VQLS cost functions, and hence VQLS cannot be efficiently simulated classically. 

\subsubsection{Ansatz} \label{sec:ansatz1}

\begin{figure}[t]
	\centering
	\includegraphics[width=\columnwidth]{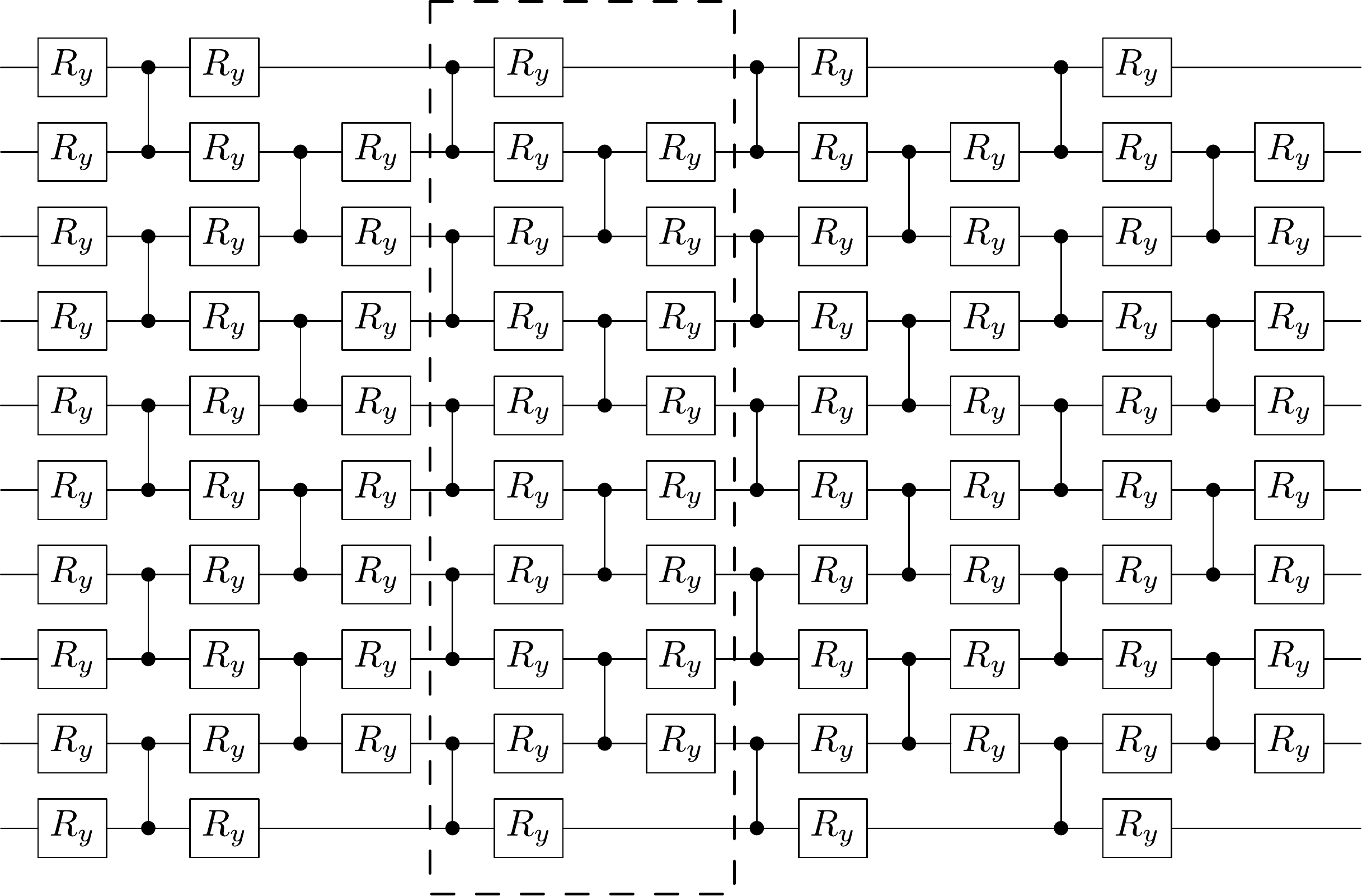}
	\caption{ Fixed-structure layered Hardware-Eficient Ansatz for $V(\vec{\alpha})$. As indicated by the dashed box, each layer is composed of controlled-$Z$ gates acting on alternating pairs of neighboring qubits which are preceded and followed by single qubit rotations around the $y$-axis, $R_y (\alpha_i)= e^{-i\alpha_i Y/2}$.  Shown is the case of four layers and $n=10$ qubits. The number of variational parameters and gates scales linearly with $n$: for $n = 50$, four layers of this ansatz consist of $640$ gates and $440$ variational parameters.
	\label{fig:ansatz}}
\end{figure}

In the VQLS algorithm, $\ket{x}$ is prepared by acting on the $\ket{\vec{0}}$ state with a trainable gate sequence $V(\vec{\alpha})$. Without loss of generality, $V(\vec{\alpha})$ can be expressed in terms of $L$ gates from a gate alphabet $\mathcal{A}=\{G_k(\alpha)\}$ as
\begin{equation} \label{eq-Vseq}
    V(\vec{\alpha})=G_{k_L}(\alpha_L)\ldots G_{k_i}(\alpha_{i})\ldots G_{k_1}(\alpha_1)\,.
\end{equation}
 Here $\vec{k}=(k_L,\ldots,k_1)$ identifies the types of gates and their placement in the circuit (i.e., on which qubit they act), while $\vec{\alpha}$ are continuous parameters. 
When working with a specific quantum hardware, it is convenient to choose a Hardware-Efficient Ansatz~\cite{kandala2017hardware}, where $\mathcal{A}$ is composed of gates native to that hardware. This reduces the gate overhead that arises when implementing the algorithm in the actual device. We use the term ``fixed-structure ansatz'' when the gate structure of $V(\vec{\alpha})$ is fixed (i.e., when $\vec{k}$  is fixed), and when one only optimizes over  $\vec{\alpha}$. Figure~\ref{fig:ansatz} shows an example of such an ansatz, with $\mathcal{A}$ composed of single qubit $y$-rotations and controlled-$Z$ gates. We employ the ansatz in Fig.~\ref{fig:ansatz} for the  heuristics in Section~\ref{sec:ising}. Let us remark that this ansatz can have trainability issues~\cite{mcclean2018barren,cerezo2020cost} for large-scale problems.

Strategies such as layer-by-layer training~\cite{grant2019initialization} and correlating the $\vec{\alpha}$ parameters~\cite{volkoff2020large}
have been shown to be effective to address these trainability issues. In addition, trainability could be further improved by combining these strategies with more advanced ansatz architectures, and we now consider two such architectures. First we discuss a ``variable structure ansatz''~\cite{cincio2018learning,VQSD}, where one optimizes over the gate angles and the gate placement in the circuit, i.e., where one optimizes over $\vec{\alpha}$ and also over $\vec{k}$. We employ such ansatz for our heuristics in  Section~\ref{sec:random}. We refer the reader to Appendix~\ref{ap:variable} for a discussion of the optimization method employed for a variable structure ansatz.

In addition to the aforementioned ansatz, one can also employ the Quantum Alternating Operator Ansatz (QAOA)~\cite{qaoa2014,nasaQAOA2019} to construct the unitary $V(\vec{\alpha})$ and avoid trainability issues. The QAOA consists of evolving the  $H^{\otimes n}\ket{\vec{0}}$ state (where $H$  denotes the Hadamard unitary) by two Hamiltonians for a specified number of layers, or rounds. These Hamiltonians are conventionally known as \textit{driver} and \textit{mixer} Hamiltonians, and respectively denoted as  $H_D$ and $H_M$. Since the ground state of both $H_G$ and $H_L$ is $\ket{x_0}$, we can either use \eqref{eqn:hamiltonian-of-global-cost} or \eqref{eqn:hamiltonian-of-local-cost} as the driver Hamiltonian $H_D$.  Evolving with $H_D$ for a time $\alpha_i$  corresponds to the unitary operator $U_D(\alpha_i) := e^{- i H_D \alpha_i}$. Moreover, one can take the mixer Hamiltonian to be the conventional $H_M = \prod_{i = 1}^{n} X_i$, where $X_i$ denotes Pauli $X$ acting on the $i$th qubit. Accordingly, evolving with $H_M$ for a time $\alpha_j$ yields the unitary operator $U_M(\alpha_j) := e^{-i H_M \alpha_j}$. The trainable ansatz $V(\vec{\alpha})$ is then obtained by alternating the unitary operators $U_D(\alpha_i)$ and $U_M(\alpha_j)$  $p$ times:
\begin{equation} \label{eqn:qaoa-ansatz}
    V(\vec{\alpha}) = e^{- i H_M \alpha_{2p}} e^{- i H_D \alpha_{2p-1}}  \cdots  e^{- i H_M \alpha_2} e^{- i H_D \alpha_1}.
\end{equation}
In this ansatz, each $\alpha_i$ is a trainable continuous parameter. 
We note that QAOA is known to be universal as the number of layers $p$ tends to infinity~\cite{qaoa2014, Lloyd_2018}, and that finite values of $p$ have obtained good results for several problems~\cite{qaoaMaxCut2018, zhou2018quantum, Crooks_2018}.  In the Supplemental Material we present results of a small scale implementation of VQLS with a QAOA ansatz.

Let us remark that Ref.~\cite{HHL} showed that it is possible to efficiently generate an accurate approximation to the true solution $\ket{x_0}$, i.e., with a number of gates that is polynomial in $n$, assuming certain constraints on $A$ and $\vec{b}$. Therefore, in principle, one may efficiently approximate these sort of solutions with a universal variational ansatz, such as the ones that we discussed above.

\subsubsection{Training algorithm}

There are several classical optimizers that may be employed to train  $V(\vec{\alpha})$ and minimize the cost functions of VQLS. For example, our heuristics in  Section~\ref{sec:ising} employ an optimization method that, at each iteration, chooses a random direction $\vec{w}$ in the parameter space along which to perform a line search, i.e., to solve $\min_{s\in \mathbb{R}} C(\vec{\alpha}+ s \vec{w})$.  On the other hand, in Section~\ref{sec:random} we perform an  optimization where all the parameters in $\vec{\alpha}$ are independently  optimized at each iteration.

In addition, there has been an increasing interest in gradient-based methods for VHQCAs~\cite{kubler2019adaptive,arrasmith2020operator,sweke2019stochastic} as it has been shown that the first-order gradient information can be directly measured~\cite{mitarai2018quantum, schuld2019evaluating} and can lead to faster rates of convergences to the optimum \cite{harrow2019low}. To enable gradient-based strategies, Appendix~\ref{ap-gradient-based} derives explicit formulas for the gradients of the cost functions and shows that the same circuits used to compute the cost functions can be used to compute their gradients. Finally, when employing the QAOA ansatz we leverage literature on QAOA-specific training (for instance, Ref.~\cite{zhou2018quantum}).

\subsubsection{Resilience to noise}

Recently, it was shown that certain VHQCAs, specifically those for compiling, can exhibit noise resilience in the sense that the optimal parameters are unaffected by certain noise models~\cite{sharma2019noise}. However, the generality of this Optimal Parameter Resilience (OPR) phenomenon is not clear. For the VQLS algorithm, the analysis of noise is made complicated by the fact that different quantum circuits are used to compute different terms in the cost function. However, perhaps surprisingly, we are able to prove that VQLS does exhibit the OPR phenomenon. Specifically, we show in Appendix~\ref{sec:Resilience} that our normalized cost functions $C_G$ and $C_L$ are both resilient to global depolarizing noise. We also show that $C_L$ is resilient to measurement noise. This is encouraging since $C_L$ is our proposed cost function of choice for large-scale implementations, although we leave an analysis of more complicated noise models for future work.


Because of OPR, we are able to train the variational ansatz and learn the correct parameters which prepare the solution $\ket{x}$. However, we still have a noisy estimate of the cost function which affects the termination conditions (\eqref{eq-costfunctionsbounds} or \eqref{eq-costfunctionsboundsTightened}) used to achieve the desired precision. In what follows, we outline an error mitigation procedure known as Probabilistic Error Cancellation (PEC)~\cite{temme2017error} which allows us to certify the termination conditions in the presence of noise.

The basic idea of PEC is to represent an ideal gate in a basis of noisy gates which a given quantum computer can implement. Exact procedures to do so are known for simple noise models such as depolarizing and amplitude damping noise. Generally, as long as the gates which a quantum computer can implement form a basis for unitary operations, an arbitrary unitary can be expressed in this basis by solving a linear program~\cite{temme2017error}. Using the notation of Ref.~\cite{temme2017error}, an ideal circuit $U_{\beta}$ can thus be written
\begin{equation}
    U_\beta = \gamma_\beta \sum_{\alpha \in \Omega_T} \sigma_{\alpha}(\beta) p_{\alpha}(\beta) \mathcal{O}_{\alpha} \,.
\end{equation}
Here, $p_\alpha(\beta)$ is a probability distribution, $\sigma_\alpha(\beta) = \pm 1$, and $\gamma_\beta$ is the negativity of the quasi-probability distribution $a_\alpha(\beta) := \gamma_\beta \sigma_\alpha (\beta) p_\alpha(\beta)$. For any observable $A^\dagger = A$, the noise-free estimate is 
\begin{align}
\nonumber
    E^*(\beta) &:= \Tr [ A \mathcal{U}_\beta (|0\rangle \langle 0| )] \\
    &= \gamma_\beta \sum_{\alpha \in \Omega_T} \sigma_\alpha (\beta) p_\alpha (\beta) \text{Tr} [ A \mathcal{O}_\alpha (|0\rangle \langle0|) ] \,. 
\end{align}
where $T$ is the number of gates in the circuit, and $\Omega_T$ is the index set over all possible circuits which has size $|\Omega_T| = m^T$, where $m$ is the number of gates in the noisy basis.
This equation is exact, but the summation has exponentially terms in the number of gates $T$. 
However, we may define an estimate of the noise-free observable as the random variable
\begin{equation} \label{eq-pec-estimate}
    \hat{E} (\beta) := \frac{\gamma_\beta}{M} \sum_{i = 1}^M \sigma_\beta (\alpha_i) \text{Tr} [ A \mathcal{O}_{\alpha_i} ( |0\rangle \langle 0| )^{\otimes n} ] \,,
\end{equation}
where $M$ is the number of samples. To achieve precision $|\hat{E} (\beta) - E^* (\beta) | \le \delta$, it is shown in~\cite{temme2017error} that one can take $M = (\gamma_\beta / \delta)^2$ samples. Concretely, this means running $M = (\gamma_\beta / \delta)^2$ circuits sampled from the distribution $p_\alpha(\beta)$ and combining the results via~\eqref{eq-pec-estimate}.

We can apply this to VQLS as follows. Let $C \in \{C_G, C_L, \hat{C}_G, \hat{C}_L\}$ be a cost function and $\tilde{C}$ be its noisy estimate. The cost $C$ is computed by evaluating $\text{poly}(L)$ circuits (terms). Using PEC to compute each term with $M = (\gamma_\beta / \delta)^2$ circuits, we have $|\tilde{C} - C| \le \poly(L) \delta $ since the error is additive. To achieve 
\begin{equation}
    |\tilde{C} - C| \le \delta 
\end{equation}
for any $\delta > 0$, we thus may run
\begin{equation}
    M = \frac{\gamma_\beta^2 \poly(L)}{\delta^2}
\end{equation}
circuits. 

In summary, OPR shows that we can obtain the optimal parameters for certain noise models, meaning that we will indeed obtain the optimal solution $\ket{x}$ to the QLSP. The above procedure based on PEC shows that, with polynomial overhead, we can verify if the solution is optimal by satisfying the termination conditions. Note that this overhead is not required during the training phase of the VQLS but rather only after the training phase has halted. If the verification fails, additional training may be done, or the training may be restarted.

\pagebreak
\raggedbottom
\subsection{Heuristic Scaling}\label{Sec-Heuristic}

\begin{figure}[t]
	\centering
	\includegraphics[width=\columnwidth]{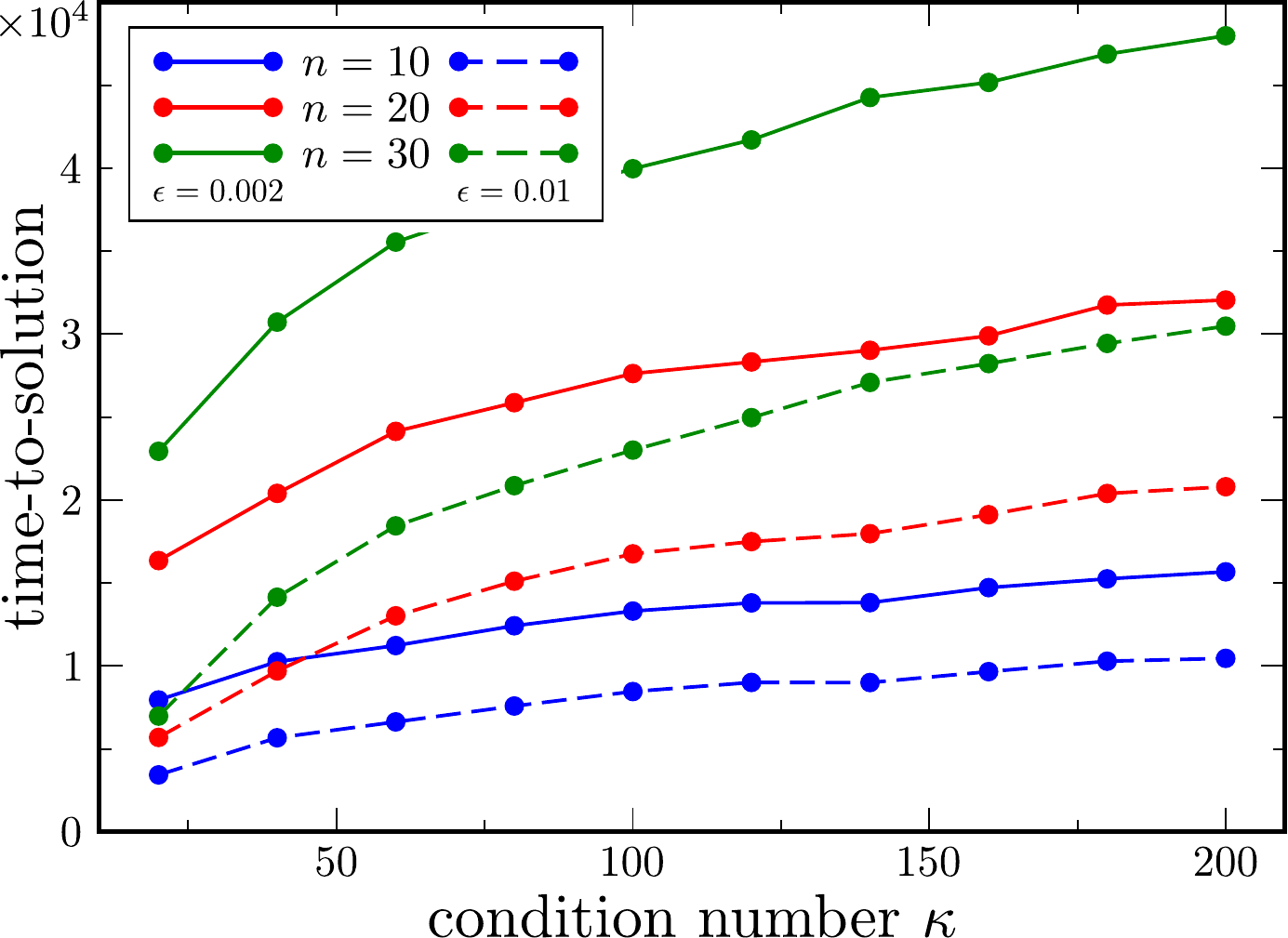}
	\caption{Scaling with $\kappa$ for the Ising-inspired QLSP. The time-to-solution is the number of executions needed to guarantee a precision  of $\epsilon=0.002$ (solid line) and $\epsilon=0.01$ (dashed line). Curves are shown for $n=10,20,30$ qubits. In each case we averaged over 30 runs of the VQLS algorithm  with four layers of the Layered Hardware-Efficient Ansatz of Fig.~\ref{fig:ansatz}, and we trained the gate sequence by minimizing $\CN_L$ of \eqref{eqn:local-cost1}. While the the $\kappa$ scaling appears to be sub-linear here, it is known that linear scaling is optimal in general~\cite{HHL}, and hence the observed scaling is likely specific to this example. \label{fig:gvskappa}}
\end{figure}

Here we study the scaling of VQLS  with the condition number $\kappa$, error tolerance $\epsilon$, and number of qubits~$n$.  First we consider a specific QLSP for which  $\ket{x_0}$ admits an efficient matrix-product-state representation,   allowing us to simulate large values of $n$. We then consider QLSPs where the matrix $A$ is randomly generated. In both cases we restrict $A$ to be a sparse matrix, which is standard for QLSPs~\cite{HHL}, and we simulate VQLS without finite sampling. Moreover, we quantify the run time of VQLS with the \textit{time-to-solution}, which refers to the number of exact cost function evaluations during the optimization needed to guarantee that $\epsilon$ is below a specified value. In practice, for large-scale implementations where the true solution $\ket{x_0}$ is unknown, $\epsilon$ cannot be directly calculated. Rather, one can use the operational meaning of our cost function in~\eqref{eq-costfunctionsbounds} to upper-bound $\epsilon$. Hence, we take this approach in all of our heuristics, i.e., we use the value of the cost, combined with \eqref{eq-costfunctionsbounds}, to determine the worst-case~$\epsilon$. We emphasize that, while it is tempting to directly compute $\epsilon$ from \eqref{eq-epsilon} in one's heuristics, this is essentially cheating since $\ket{x_0}$ is unknown, and this is why our certification procedure is so important.

\begin{figure}[t]
    \centering
    \includegraphics[width=\columnwidth]{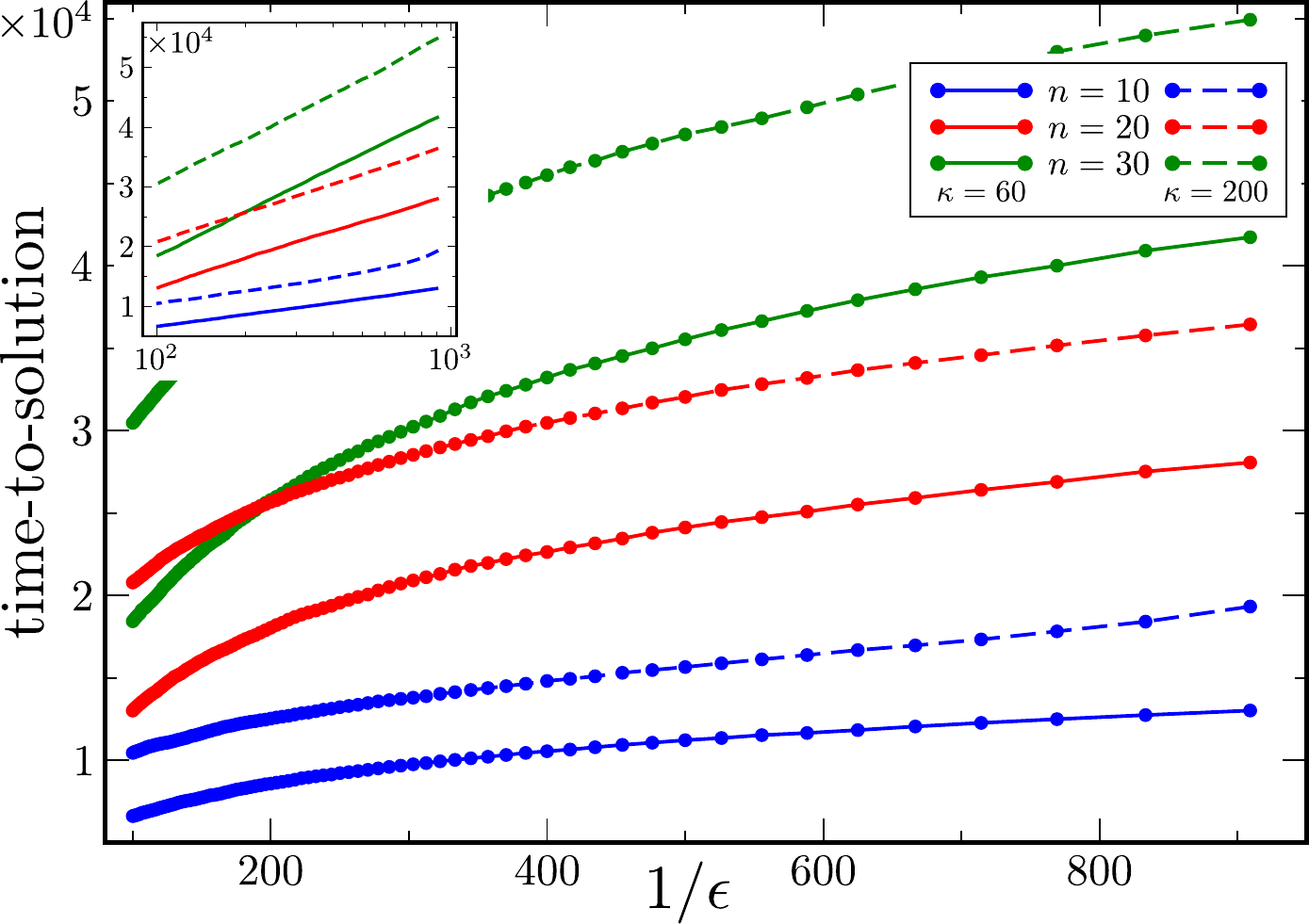}
    \caption{Scaling with $1/\epsilon$ for the Ising-inspired QLSP. Curves are shown for $n=10,20,30$ qubits, with $\kappa=60$ (solid line) or $\kappa=200$ (dashed line). In all cases $V(\vec{\alpha})$ was composed of four layers of the Layered Hardware-Efficient Ansatz of Fig.~\ref{fig:ansatz}, and we trained the local cost $\CN_L$. The time-to-solution was obtained by averaging over 30 runs of the VQLS algorithm. The inset depicts  the same data in a logarithmic scale. The dependence on $1/\epsilon$ appears to be logarithmic, i.e., linear on a logarithmic scale.}
    \label{fig:EpsilonDependence}
\end{figure}

\makeatletter
\afterpage{\global\let\@textbottom\relax \global\let\@texttop\relax}

\subsubsection{Ising-inspired QLSP}\label{sec:ising}

Here we numerically simulate VQLS to solve the QLSP defined by the sparse matrix
\begin{equation}\label{eqn:ExampleQLSP}
\begin{aligned} 
    A&=\frac{1}{\zeta}\left(\sum_{j=1}^{n}X_j+J\sum_{j=1}^{n-1}Z_jZ_{j+1} +\eta \id\right)\,,\\
    \ket{b}&=H^{\otimes n}\ket{\vec{0}}\,,
\end{aligned}
\end{equation}
where the subscripts in \eqref{eqn:ExampleQLSP} denote the qubits acted upon non-trivially by the Pauli operator. Here, we set $J=0.1$. The parameters $\zeta$ and $\eta$ are chosen such that the smallest eigenvalue of $A$ is $1/\kappa$ and its largest eigenvalue is $1$, which involves analytically computing~\cite{he2017boundary} the smallest eigenvalue of the first two terms of $A$ and then re-scaling $A$. As previously mentioned, this QLSP example is motivated from the fact that for $J=0$ the solution is given by $\ket{x_0}=\ket{b}$. Hence for small $J$, $\ket{x_0}$ admits an efficient matrix-product-state representation.

\textit{Dependence on $\kappa$}: Figure~\ref{fig:gvskappa} shows our results, plotting time-to-solution versus $\kappa$ for the  QLSP in~\eqref{eqn:ExampleQLSP}.
Our numerical results were obtained by employing the
layered Hardware-Efficient Ansatz of Fig.~\ref{fig:ansatz}, and by training the local cost $\CN_L$ for different values of $n$.
Figure~\ref{fig:gvskappa} shows that as the condition number $\kappa$ is increased, the time-to-solution needed to achieve a given $\epsilon$ increases with a scaling that appears to be sub-linear. Hence VQLS scales efficiently with $\kappa$ for this example. It is known that linear scaling is optimal~\cite{HHL}. Hence we expect that the scaling observed here is specific to this example, and indeed the example in the next subsection shows scaling that is closer to linear.

\textit{Dependence on $\epsilon$}: To study the scaling of VQLS with~$\epsilon$, we numerically solved the QLSP in~\eqref{eqn:ExampleQLSP} for different values of $\kappa$ and~$n$. In all cases we trained the gate parameters by optimizing the $\CN_L$ cost function.  Figure~\ref{fig:EpsilonDependence} shows the time-to-solution versus $1/\epsilon$. These results show that as $1/\epsilon$ grows, the time-to-solution exhibits a logarithmic growth.

\begin{figure}[t]
    \centering
    \includegraphics[width=\columnwidth]{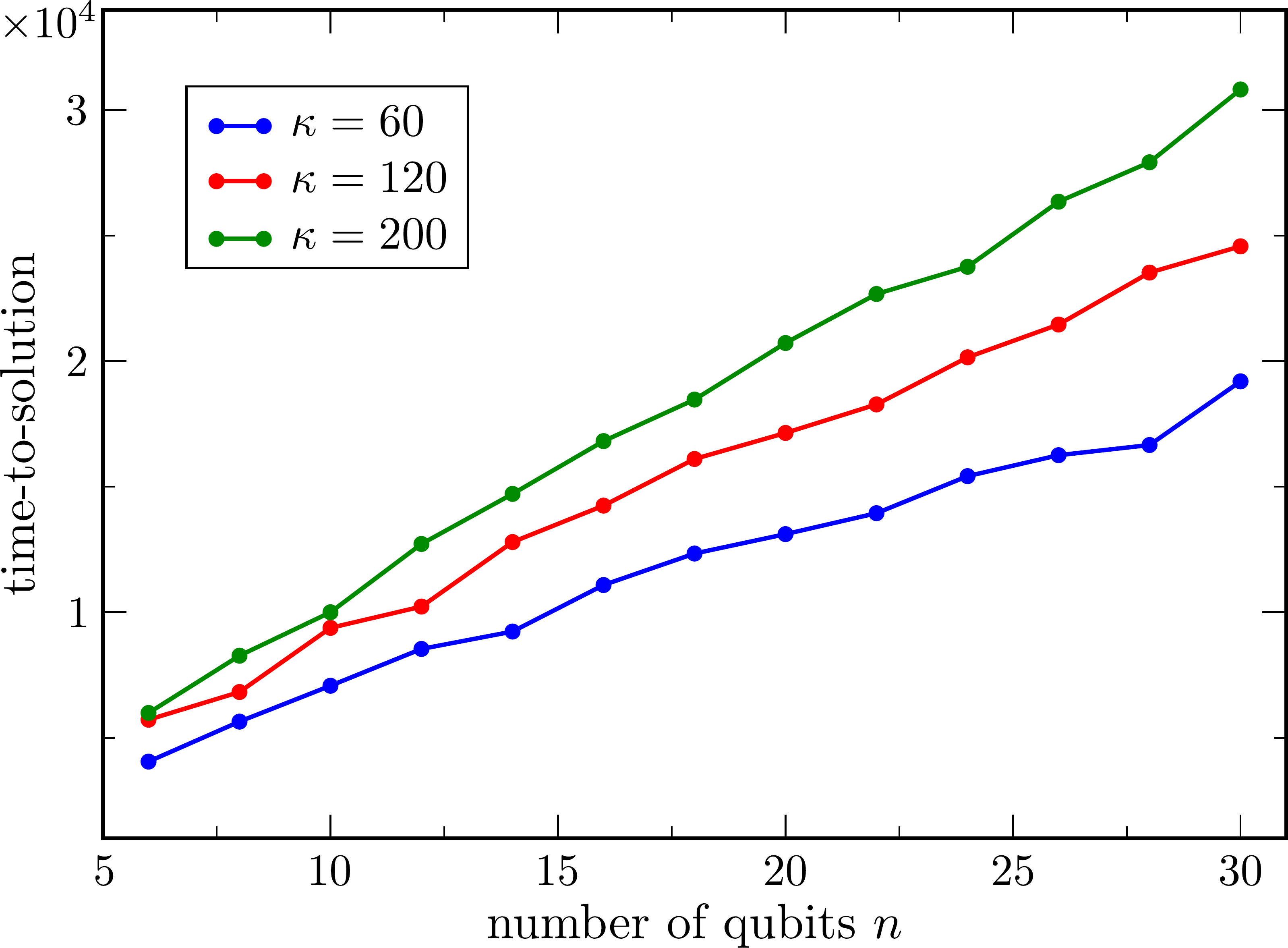}
    \caption{Scaling with $n$ for the Ising-inspired QLSP. Curves are shown for $\epsilon=0.01$ and for $\kappa=60,120,200$. In all cases we trained the local cost $\CN_L$ with four layers of the Layered Hardware-Efficient Ansatz of Fig.~\ref{fig:ansatz}. The dependence on $n$ appears to be linear (logarithmic in $N$) for this example.}
    \label{fig:nDependence}
\end{figure}

\textit{Dependence on $n$}: The QLSP of \eqref{eqn:ExampleQLSP} allows us to increase the number of qubits and analyze the scaling of VQLS with $n$. Here we implemented VQLS with $n=6,8,\ldots,30$ and for  $\kappa=60,120,200$ by training the local cost function $\CN_L$.  Figure~\ref{fig:nDependence} shows  time-to-solution versus $n$. As the number of qubits increases the time-to-solution needed to guarantee a particular $\epsilon$ with $\kappa$ fixed appears to increase linearly with $n$. This corresponds to logarithmic scaling in the linear system size $N$, analogous to that of the HHL algorithm~\cite{HHL}.

\subsubsection{Randomly generated QLSP}\label{sec:random}

In this section we present scaling results for the case when the  matrix $A$ is randomly generated with the form
\begin{equation} \label{eq:randA}
A = \xi_1\Big(\id + \xi_2\sum_{j} \sum_{k \neq j } p a_{j,k} \sigma^\alpha_j \sigma^\beta_k \Big)\,.
\end{equation}
Here  $p$ is either $0$ or $1$ according to a fixed binomial distribution,  $a_{j,k}$ are random weights in $(-1,1)$, and $\sigma^\alpha_j$ is the Pauli matrix acting on qubit $j$ with $\alpha = x,y,z$. For each $j,k=1,\ldots,n$ in \eqref{eq:randA}, $\alpha$ and $\beta$ are randomly chosen. Finally, we remark that $\xi_1$, and $\xi_2$ are normalization coefficients that rescale the matrix so that its largest eigenvalue is 1 and its smallest is $1/\kappa$ (where $\kappa$ is fixed).

For a given number of qubits $n$, we randomly created a matrix $A$ according to \eqref{eq:randA},  and we  ran four independent instances of VQLS. We then selected the best run, i.e., the instance that required the smallest number of cost function evaluations to reach a specified value of guaranteed~$\epsilon$ (guarenteed via~\eqref{eq-costfunctionsbounds}). This procedure was then repeated for 10 independent random matrices $A$, and the  time-to-solution was obtained as the average of the best run for each matrix.

\begin{figure*}[t]
    \centering
    \includegraphics[width=\linewidth]{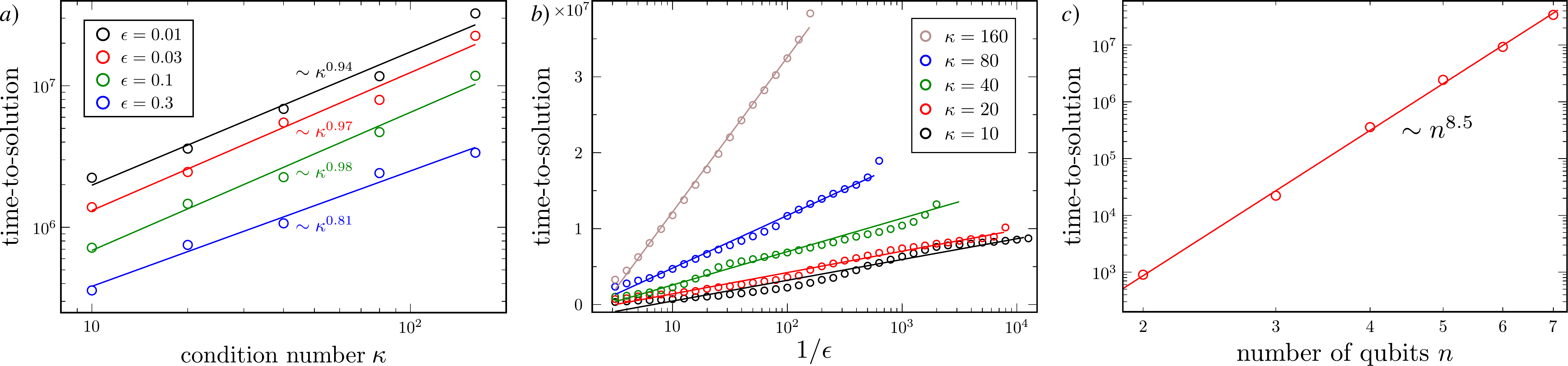}
    \caption{VQLS heuristic scaling for random matrices generated according to~\eqref{eq:randA}. The time-to-solution is the number of executions needed to guarantee a desired precision $\epsilon$. In all cases we employed a variable-structure ansatz $V(\vec{\alpha})$ as described in Appendix~\ref{ap:variable}, and we trained the local cost $C_L$ of~\eqref{eqn:local-cost1}.  a) Time-to-solution versus $\kappa$ for a system of $n=4$ qubits. Axes are shown in a log-log scale. For each value of $\epsilon$ the data were fitted with a power function $\kappa^m$ and in all cases $m<1$, suggesting that the  $\kappa$ scaling appears to be sub-linear. b) Time-to-solution versus $1/\epsilon$ for a system of $n=4$ qubits.  The $x$ axis is shown in a log scale.  Each curve corresponds to a different condition number. For all values of $\kappa$ the data were fitted with a linear function, implying that the $1/\epsilon$ scaling is logarithmic. c) Time-to-solution versus $n$  needed to guarantee $\epsilon=0.3$. All matrices   had a condition number $\kappa=10$. The plot employs a log-log scale. The data were fitted with a power function $y\sim n^{8.5}$, suggesting that the $N$ dependence is polylogarithmic.  }
    \label{fig:randomplots}
\end{figure*}

\textit{Dependence on $\kappa$}: In Fig.~\ref{fig:randomplots}(a) we show the time-to-solution versus $\kappa$ for matrices  randomly generated  according to~\eqref{eq:randA}, and for $n=4$. Here we employed a variable-structure ansatz as described in Appendix~\ref{ap:variable}, and we trained the local cost in~\eqref{eqn:local-cost1}. Different curves represent different desired precision $\epsilon$. The data were plotted in a log-log scale and each curve was fitted with a power function $\kappa^m$. In all cases we found $m<1$, indicating that the scaling in $\kappa$ for these examples is at worst linear. Linear scaling in $\kappa$ is known to be optimal~\cite{HHL}.

\textit{Dependence on $\epsilon$}: Let us now analyze the  scaling of VQLS with respect to $\epsilon$ for matrices with different condition numbers. Figure~\ref{fig:randomplots}(b) depicts the time-to-solution versus $\epsilon$ for matrices  randomly generated  according to~\eqref{eq:randA}, and for $n=4$. 
All curves were fitted with a linear function, and since the $x$ axis is in a logarithmic scale, the dependence on $1/\epsilon$ appears to be logarithmic. Upon examining Figures~\ref{fig:randomplots}(a) and~\ref{fig:randomplots}(b) collectively, the apparent scaling of the time-to-solution with $\kappa$ and $\log(1/\epsilon)$ seems to exhibit a multiplicative behavior.

\textit{Dependence on $n$}: In Fig.~\ref{fig:randomplots}(c) we present the time-to-solution versus $n$ needed to guarantee  $\epsilon=0.3$ for QLSPs with $n=2,\ldots,7$. All  matrices $A$ had condition number $\kappa=10$. The data were  fitted with  a power function and we obtained the relation $y\sim n^{8.5}$. This corresponds to polylogarithmic scaling in $N$, which is the standard goal of quantum algorithms for the QLSP~\cite{CKS, Ambainis, Chakraborty, Yigit}.

We refer the reader to the Supplemental Material for additional numerical simulations of VQLS for other QLSP examples, both with a Hardware-Efficient Ansatz and with a QAOA ansatz. These examples also exhibit efficient scaling behavior.

\subsection{Implementation on quantum hardware}

Here we present the results of a $1024 \times 1024$  (i.e., 10-qubit) implementation of VQLS using Rigetti's \textit{16Q Aspen-4} quantum computer. Specifically, we solved the QLSP defined by the matrix $A$ in~\eqref{eqn:ExampleQLSP}, with $\zeta = \eta = 1$, and where the vector $|b\rangle = |0\rangle$ was the all zero state. The ansatz consisted of $R_y(\alpha_i)$ gates acting on each qubit. To adapt to hardware constraints, we computed the cost function $C_G$ in \eqref{eqn:global-cost2} by expanding the effective Hamiltonian $H_G$ in terms of Pauli operators and then employing Rigetti's quantum computer to estimate the expectation values of these terms.

\begin{figure}[t]
	\centering
	\includegraphics[width=.9\columnwidth]{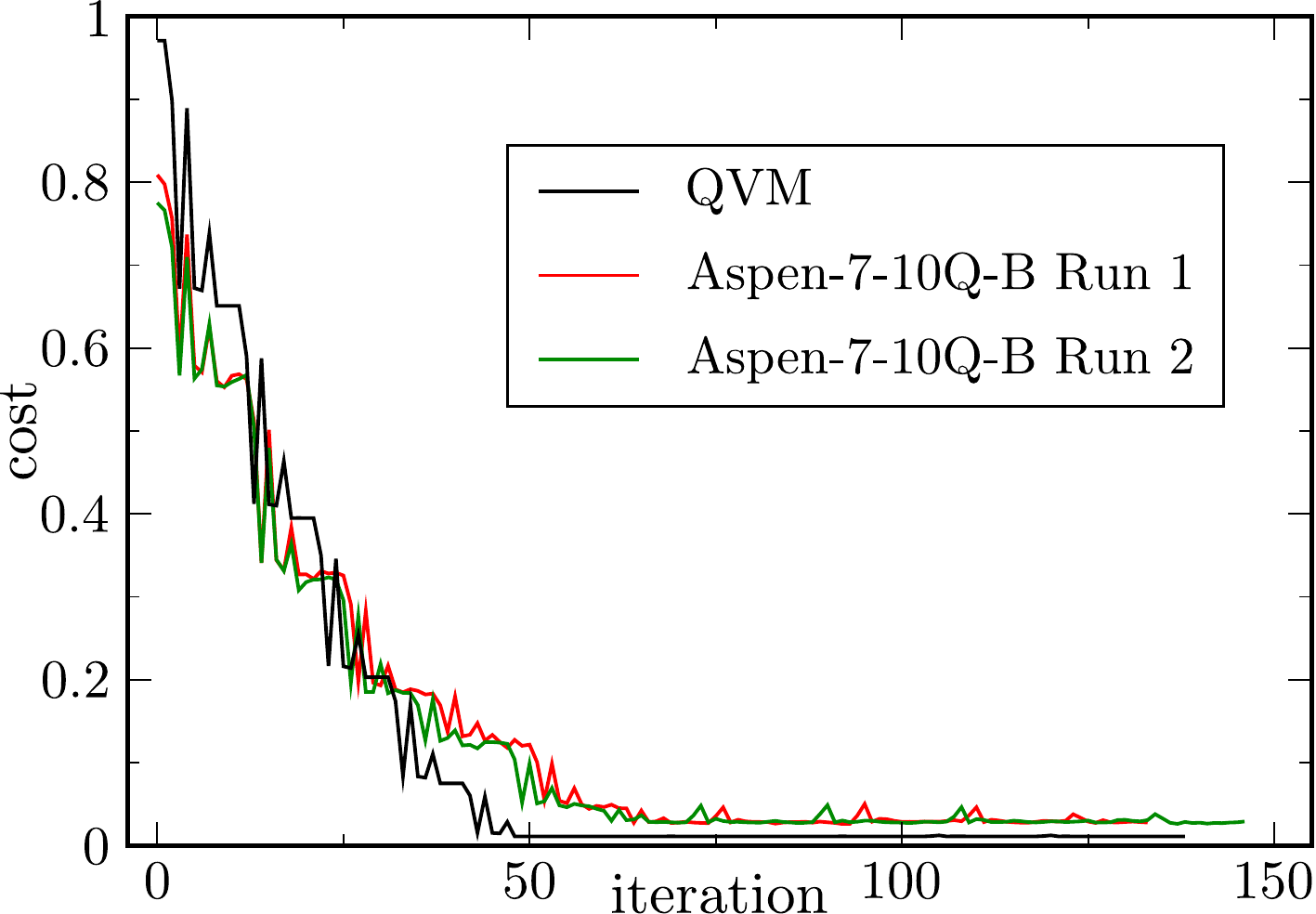}
	\caption{ Implementation of VQLS on Rigetti's quantum hardware. Cost function $\CN_G$ is plotted against the number of optimization steps, where $A$ is defined in~\eqref{eqn:ExampleQLSP}.  One can observe that for each QPU run the cost function is reduced to a value below $10^{-1}$. Due to noise present in the quantum device the cost does not go to zero. \label{fig:10q_rigetti}}
\end{figure}

The results of two representative VQLS runs are shown in Fig.~\ref{fig:10q_rigetti}. As shown, the cost function data obtained by training in a  quantum computer closely matches the one obtained from training on a  noiseless simulator. For each run on the QPU, the value of the cost function approaches zero, indicating that a good solution to the linear system was found. 

Additional experiments performed on quantum hardware are presented in the Supplemental Material.

\section{Discussion}

In this work, we presented a variational quantum-classical algorithm called VQLS for solving the quantum linear systems problem. On the analytical side, we presented four different faithful cost functions, we derived efficient quantum circuits to estimate them while showing that they are difficult to estimate classically, and we proved operational meanings for them as upper bounds on $\epsilon^2 / \kappa^2$. On the numerical side, we studied the scaling of the VQLS run time by solving non-trivial problems of size up to $2^{50}\times2^{50}$. For the examples considered, we found VQLS to scale efficiently, namely, at worst linearly in $\kappa$, logarithmically in $1/\epsilon$, and polylogarithmically  in the linear system size $N$. 

It remains to be seen how the VQLS training is affected by finite sampling, which is not accounted for in our heuristics. Our solution verification procedure in Sec.~\ref{sec:OpMeaning} will require the shot noise to appropriately scale with $\epsilon$ and $\kappa$ as dictated by \eqref{eq-costfunctionsbounds}. Namely, the number of shots would need to scale as $(\kappa/\epsilon)^4$, although this complexity might be reduced if one does not require solution verification. 

Furthermore, we utilized Rigetti's Quantum Cloud Services to implement VQLS for a particular problem up to a size of $1024 \times 1024$, which to our knowledge is the largest implementation of a linear system on quantum hardware. Interestingly, with our implementation on Rigetti's hardware, we noticed some preliminary evidence of noise resilience, along the same lines as those discussed in Ref.~\cite{sharma2019noise} for a different variational algorithm. Namely, we noticed optimal parameter resilience, where VQLS learned the correct optimal parameters despite various noise sources (e.g., measurement noise, decoherence, gate infidelity) acting during the cost evaluation circuit. We will explore this in future work, including tightening our certification bound in \eqref{eq-costfunctionsbounds} when accounting for noise.

Finally, we discuss how VQLS fits into the larger literature on quantum algorithms for linear systems. Most prior algorithms rely on time evolutions with the matrix $A$~\cite{HHL,Ambainis,CKS} or a simple function of it~\cite{Yigit}. In these algorithms, the duration of the time evolution is $O(\kappa)$ in order to prepare a state $\ket{x}$ that is $\epsilon$-close to the correct answer. In general, this can only be achieved with a quantum circuit of size linear in $\kappa$ as per the ``no fast-forwarding theorem''~\cite{berry2007efficient,atia2017fast}. This is even true if there exists a very short quantum circuit that prepares the desired state $\ket{x}$. The non-variational algorithms simply cannot exploit this fact. On the other hand, a variational algorithm with a short-depth ansatz might be used to prepare such a state.

This does not mean, however, that the overall complexity of the variational algorithm does not depend on the condition number. This dependence enters through the stopping criteria given in \eqref{eq-costfunctionsbounds}. As the condition number increases, the cost has to be lowered further in order to guarantee an error of $\epsilon$. This will undoubtedly require more iterations of the variational loop to achieve. In effect, our variational approach trades the gate complexity of non-variational algorithms with the number of iterations for a fixed circuit depth. This trade-off can be useful in utilizing NISQ devices without error correction.

We remark that other variational approaches to the QLSP distinct from ours were very recently proposed~\cite{Xiaosi,huang2019near}. Relatively speaking, the distinct aspects of our work include: (1) our quantitative certification procedure for the solution, (2) our clear approach to improve trainability for large-scale problems, (3) our novel circuits for efficient cost evalutaion, (4) our large-scale heuristics demonstrating efficient scaling, and (5) our large-scale implementations on quantum hardware. Finally, it exciting that, shortly after our paper was posted, two independent tutorials for the VQLS algorithm were created and added to IBM's open-source Qiskit textbook \cite{Qiskit-Textbook}, and to Xanadu's PennyLane library \cite{PennyLane-Tutorial}.

\appendix

\bigskip

\section{Sparse Matrices}
\label{sec:sparse}

In this section we describe how sparse matrices can be expressed as a linear combination of unitaries. 
Once this is achieved, the cost function can be computed using the methods described elsewhere in this paper.  
The construction below from Ref.~\cite{Yigit} uses a version of Szegedy walks that applies to Hermitian matrices~\cite{Sze04,BCK15}. 
Let $A$ be a $d$-sparse matrix of dimension $N$ and $n=\log_2 N$. 
We define unitary operations $U_x$, $U_y$, and $S$ that act as follows:
\begin{widetext} 
\begin{align}
    U_x \ket{j} \ket{0} \ket{0} \ket{0} &= \frac{1}{\sqrt{d}}\sum_{i\in F_j} \ket{j} \ket{i} \ket{0} \left( \sqrt{A_{ji}^*} \ket{0} + \sqrt{1-\vert A_{ji} \vert} \ket{1}\right) \;, \\
    U_y \ket{0} \ket{j'} \ket{0} \ket{0} &= \frac{1}{\sqrt{d}}\sum_{i'\in F_{j'}} \ket{i'} \ket{j'}  \left( \sqrt{A_{i'j'}} \ket{0} + \sqrt{1-\vert A_{i'j'} \vert} \ket{1}\right)\ket{0} \;, \\
    S \ket{j}\ket{j'}\ket{\cdot}\ket{\cdot} &= \ket{j'}\ket{j}\ket{\cdot}\ket{\cdot} \; .
\end{align}
\end{widetext}
The first two registers have $n$ qubits each,
the last two registers have a single qubit each, and $F_j$ is the set of indices $i$ for which $A_{ji}$ is nonzero.
It follows that
\begin{align}
\nonumber
    A \otimes \ket{\tilde 0}\!\bra{\tilde 0} &= d \; \ket{\tilde{0}} \!\bra{\tilde{0}} U_x^\dagger U_y S \ket{\tilde{0}}\!\bra{\tilde{0}} \\
    \label{eq:LCUofA}
    &= \frac{d}{4}\left( \id - e^{i\pi P }\right) U_x^\dagger U_y S \left( \id - e^{i\pi P }\right) \; ,
\end{align}
where we defined $\ket{\tilde{0}} := \ket{0}\ket{0}\ket{0}$ for the state of the last three registers and $P= \ket{\tilde 0}\!\bra{\tilde 0}$.
Equation~\eqref{eq:LCUofA} is a decomposition of $A$ as a linear combination of 4 unitaries with equal weights of $d/4$.

We assume access to an oracle for $A$ that acts as
\begin{align}
     \ket{j}\ket{i}\ket{z} & \rightarrow \ket{j}\ket{i}\ket{z\oplus A_{ji}} \;,\\
      \ket{j}\ket{l} & \rightarrow \ket{j}\ket{f(j,l)} \;.
\end{align}
Here, $j$ and $i$ label the row and column of $A$,
respectively, so that $j,i \in\{1,\ldots,N\}$, and $f(j,l)$ is the column index of the $l$'th nonzero element of $A$ in row $j$. We refer to this oracle as $\mathcal{O}_A$.  
This is the same as that used in previous works for the QLSP and Hamiltonian simulation
such as Refs.~\cite{CKS,BCK15}.

$U_x$ can then be implemented in five steps as follows:
\begin{align}
     & \ket{j}\ket{0}\ket{0}\ket{0} \ket{0}\notag\nonumber\\ &\xrightarrow[\text{Hadamards}]{\log(d)} \frac{1}{\sqrt{d}}\sum_{l=0}^{d-1} \ket{j}\ket{l}\ket{0}\ket{0} \ket{0}\nonumber\\
    &\xrightarrow[]{\mathcal{O}_{A}} \frac{1}{\sqrt{d}} \sum_{i\in F_j} \ket{j}\ket{i}\ket{0}\ket{0}\ket{0}\nonumber \\
     &\xrightarrow[]{\mathcal{O}_{A}} \frac{1}{\sqrt{d}} \sum_{i\in F_j} \ket{j}\ket{i}\ket{A_{ji}}\ket{0}\ket{0}\nonumber \\
     &\xrightarrow[]{~~M~~} \frac{1}{\sqrt{d}} \sum_{i\in F_j} \ket{j}\ket{i}\ket{A_{ji}}\ket{0}\left( \sqrt{A_{ji}^*}\ket{0} + \sqrt{1-\vert A_{ji}\vert}\ket{1} \right) \nonumber \\
    &\xrightarrow[]{\mathcal{O}_{A}} \frac{1}{\sqrt{d}} \sum_{i\in F_j} \ket{j}\ket{i}\ket{0} \ket{0}\left( \sqrt{A_{ji}^*}\ket{0} + \sqrt{1-\vert A_{ji}\vert}\ket{1} \right) \nonumber \;.
\end{align}
The third register is used to temporarily store the matrix elements of $A$ and is discarded at the end.
Its size depends on the precision with which the matrix elements of $A$ are specified. A similar procedure can be followed to implement $U_y$.

Next, we briefly analyze the gate complexity of the unitaries in the decomposition of $A$ given by Eq.~\eqref{eq:LCUofA}. 
Since both $U_x$ and $U_y$ use 3 queries to $\mathcal{O}_A$, each unitary uses 6 queries. 
Other than the queries the main gate complexity comes from the implementation of the operations $U_x$ and $U_y$. The gate complexity of both of these strongly depends on the form and precision with which the matrix elements of $A$ are specified. The gate complexity of the unitaries $e^{i\pi P}$ is $O(\log N)=O(n)$.

Finally, we note that the decomposition in Eq.~\eqref{eq:LCUofA} has a prefactor proportional to $d$ multiplying all the unitaries in the linear combination. This implies that in order to compute the cost function with a given precision, we have to estimate the expectation value of each unitary with a precision that is inversely proportional to $d$. For this reason we can only use this approach for sparse matrices where $d\ll N$.

\section{Faithfulness of the cost functions}\label{sec:faith}

We now prove \eqref{eqnCostEquivalence}, which is restated here:
\begin{align}
\CU_L \leq \CU_G \leq n \CU_L\,,\qquad
\CN_L \leq \CN_G \leq n \CN_L\,.
\end{align}
For the lower bound, let $\Pi_G = \id - \dya{\vec{0}}$ and $\Pi_L = \id - \frac{1}{n}\sum_{j=1}^n \dya{0_j} \otimes \id_{\overline{j}}$. Using the fact that $\dya{0_j} \otimes \id_{\overline{j}}\geq \dya{\vec{0}}$, we have $\Pi_G \geq \Pi_L$ and hence $H_G \geq H_L$. This implies that $\CU_G\geq \CU_L$ and $\CN_G\geq \CN_L$. 

For the upper bound, note that $\Pi_G = \sum_{\vec{z}\neq \vec{0}} \dya{\vec{z}}$. Let $\SC_j$ denote the set of all bitstrings that have a one at position $j$, and let $\SC = \bigcup_j \SC_j$ denote the union of all of these sets. Then 
\begin{align}
n\Pi_L = \sum_j \sum_{\vec{z}\in \SC_j} \dya{\vec{z}} \geq \sum_{\vec{z}\in \SC} \dya{\vec{z}} = \sum_{\vec{z}\neq \vec{0}} \dya{\vec{z}} = \Pi_G\,.    
\end{align}
Hence we have $n H_L \geq H_G$, which implies $n\CU_L \geq \CU_G$ and $n\CN_L \geq \CN_G$.

Equation~\eqref{eqnCostEquivalence} implies the faithfulness of the cost functions as follows. Because $\Pi_G \geq 0$ and $\Pi_L \geq 0$, we have that all four cost functions are non-negative. Furthermore, it is clear that if $A\ket{x} =\ket{b}$, then we have $\CU_G = \CN_G = \CU_L = \CN_L = 0$. Conversely, assuming that $A\ket{x} \neq \ket{b}$ implies that $C_G > 0$ and hence that all four cost functions are positive. Therefore, all four cost functions are faithful, vanishing if and only if $A\ket{x} =\ket{b}$.

\begin{figure*}[t]
	\centering
	\includegraphics[width=2\columnwidth]{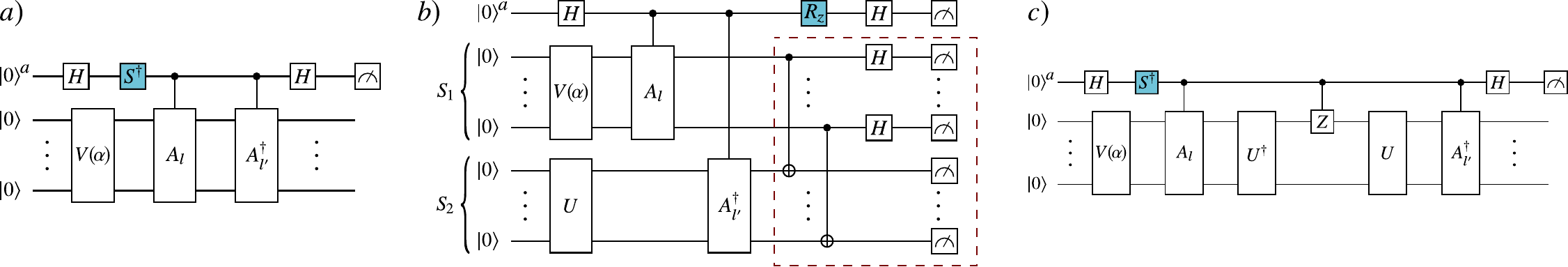}
	\caption{a) Circuit for the Hadamard Test used to compute the coefficients $\beta_{ll'}=\mte{\vec{0}}{V\ad A_{l'}\ad A_{l} V }$ and calculate the inner product $\ip{\psi}{\psi}$ of \eqref{eq-psipsi}. The phase gate in the colored box is excluded when calculating the real part of $\beta_{ll'}$ and included when calculating its imaginary part. b) Hadamard-Overlap Test used to compute the coefficients $\gamma_{ll'}$ defined in \eqref{eq-gamall}. The Overlap circuit of Refs.~\cite{garcia2013swap, cincio2018learning} is indicated in the dashed box.  Here, the $R_z$ gate in the colored box denotes a rotation about the $z$ axis of an angle~$-\pi/2$. Excluding (including) this rotation allows one to calculate the real (imaginary) part of $\gamma_{ll'}$. As explained in the text, additional post-processing is required.  c) Hadamard Test circuit for computing $\delta_{ll'}^{(j)}$ as defined in \eqref{eq-deltall2}. Shown here is case when $j=1$.} \label{fig:hadamardtest}
\end{figure*}

\section{Cost evaluation circuits} \label{Sec:cost-circuite}

In this section we present short-depth circuits for computing the cost functions of Eqs.~\eqref{eqn:global-cost1}, \eqref{eqn:global-cost2} and \eqref{eqn:local-cost1}. In particular, we introduce the Hadamard-Overlap Test circuit, which should be of interest on its own as it is likely to have applications outside of the scope of VQLS.

\subsection{Hadamard Test}

Figure~\ref{fig:hadamardtest}(a) shows a Hadamard Test which can be used to measure the coefficients $\beta_{ll'}$ defined in \eqref{eq-betall}, and used to compute $\ip{\psi}{\psi}$ as in \eqref{eq-psipsi}. When the phase gate is excluded, the probability of measuring the ancilla qubit in the $\ket{0}^a$ state is $P(0)=(1+\Re[\beta_{ll'}])/2$, while the probability of measuring it in the  $\ket{1}^a$ state is $P(1)=(1-\Re[\beta_{ll'}])/2$. Hence, by means of the Hadamard Test we can compute the real part of $\beta_{ll'}$ as  
\begin{equation}
\Re[\beta_{ll'}]=P(0)-P(1)\,.    
\end{equation}
With a similar argument it can be easily shown that by including the phase gate one can compute $\Im[\beta_{ll'}]$.

As we now show, in order to compute the coefficients $\delta_{ll'}^{(j)}$ in \eqref{eq-deltall} we can use the previous result combined with those obtained by means of the Hadamard test of Figure~\ref{fig:hadamardtest}(c). In particular, since $\dya{0_j}=(\id_j+Z_j)/2$, then we can express
\begin{equation} \label{eq-deltall2}
    \delta_{ll'}^{(j)}= \beta_{ll'} + \mte{\vec{0}}{V\ad A_{l'}\ad U (Z_j \otimes \id_{\overline{j}} ) U\ad A_l V}\,.
\end{equation}
Hence, in order to calculate $\delta_{ll'}^{(j)}$ one only needs to measure the real and imaginary parts of the matrix elements $\mte{\vec{0}}{V\ad A_{l'}\ad U (Z_j \otimes \id_{\overline{j}} ) U\ad A_l V}$, which can be accomplished by means of the circuit in Fig.~\ref{fig:hadamardtest}(c).

\subsection{Hadamard-Overlap Test}

Consider the circuit in Fig.~\ref{fig:hadamardtest}(b), which we refer to as the Hadamard-Overlap Test. A nice feature of the Hadamard-Overlap Test is that it only requires one application of both $U$ and $V$, and these unitaries do not need to be controlled, in contrast to the Hadamard Test. As explained below, the circuit for the Hadamard-Overlap Test can be obtained by combining the Hadamard Test with the Overlap circuit of Refs.~\cite{garcia2013swap, cincio2018learning}. This circuit requires $2n+1$ qubits and classical post-processing (which scales linearly with $n$) similar to that of the Overlap circuit.

Given that the Hadamard-Overlap Test combines two different ideas, let us describe its inner workings in detail. First, the circuit is initialized to the state $\ket{0}^a\otimes \ket{\vec{0}}\otimes \ket{\vec{0}}$, with $\ket{\vec{0}}=\ket{0}^{\otimes n}$. We will denote the first qubit, initialized to $\ket{0}^a$, as the ancilla, while we will refer to the next two sets of $n$-qubits as subsystems $S_1$ and $S_2$, respectively,  as in Fig.~\ref{fig:hadamardtest}(b). One then applies a Hadamard gate to the ancilla qubit,  $V$ to qubits in $S_1$, and the unitary $U$ to the qubits in $S_2$, producing the state
\begin{align}
    \ket{0}^a\otimes \ket{\vec{0}}\otimes \ket{\vec{0}}\rightarrow  \ket{+}\otimes V\ket{\vec{0}}\otimes U\ket{\vec{0}}\,.\nonumber
\end{align}
The application of the controlled $A_l$ and $A_{l'}\ad$ unitaries leads to 
\begin{align}
    &\ket{+}\otimes V\ket{\vec{0}}\otimes U\ket{\vec{0}}\nonumber\\
    &\rightarrow  \frac{1}{\sqrt{2}}\left(\ket{0}\otimes V\ket{\vec{0}}\otimes U\ket{\vec{0}}+ \ket{1}\otimes A_lV\ket{\vec{0}}\otimes A_{l'}\ad U\ket{\vec{0}}\right)\,.\nonumber
\end{align}
Next, let us consider the case when the $R_z$ gate in Fig.~\ref{fig:hadamardtest}(b) is excluded. The next Hadamard gate on the ancilla leads to 
\begin{align}
    &  \frac{1}{\sqrt{2}}\left(\ket{0}\otimes V\ket{\vec{0}}\otimes U\ket{\vec{0}} +\ket{1}\otimes A_lV\ket{\vec{0}}\otimes A_{l'}\ad U\ket{\vec{0}}\right)\nonumber\\
    &\rightarrow\ket{\psi}=
    \frac{1}{2}\ket{0}\otimes\left( V\ket{\vec{0}}\otimes U\ket{\vec{0}} + A_lV\ket{\vec{0}}\otimes A_{l'}\ad U\ket{\vec{0}}\right)\nonumber\\
    &\quad\quad\quad+ \frac{1}{2}\ket{1}\otimes\left( V\ket{\vec{0}}\otimes U\ket{\vec{0}} - A_lV\ket{\vec{0}}\otimes A_{l'}\ad U\ket{\vec{0}}\right)\,.\nonumber
\end{align}

From here, one needs to perform a measurement over state $\ket{\psi}$. As shown in Fig.~\ref{fig:hadamardtest}(b), we measure the ancilla on the computational basis and the $S_1$ and $S_2$ qubits on the Bell basis. More specifically, we can see that we are measuring the first qubit on $S_1$ and the first qubit in $S_2$ in the Bell basis, as well as the second qubit on $S_1$ and the second qubit in $S_2$ in the Bell basis, and so-on. At this point, we find it convenient to recall that the Bell basis on two qubits is composed of four states:
\begin{align}
    \ket{\phi^{00}}&=\frac{1}{\sqrt{2}}(\ket{00}+\ket{11})\,,\quad 
    \ket{\phi^{01}}&=\frac{1}{\sqrt{2}}(\ket{00}-\ket{11})\nonumber\\
    \ket{\phi^{10}}&=\frac{1}{\sqrt{2}}(\ket{01}+\ket{10})\,,\quad
    \ket{\phi^{11}}&=\frac{1}{\sqrt{2}}(\ket{01}-\ket{10})\nonumber\,,
\end{align}
Note that given a two-qubit state $\rho$, and denoting as $P(\phi^{pq})=\Tr[\rho\dya{\phi^{pq}}]$ the probability of obtaining the outcome $\ket{\phi^{peq}}$, by measuring $\rho$ on the Bell basis, then
\footnotesize
\begin{align}
 \sum_{pq=0,1}(-1)^{\delta_{pq,11}}P(\phi^{pq})
    &=P(\phi^{00})+P(\phi^{01})+P(\phi^{10})-P(\phi^{11})\nonumber\\
    &=\Tr[\rho {\rm SWAP}]\,,
\end{align}
\normalsize
where here ${\rm SWAP}$ denotes the standard two-qubit SWAP gate. The previous result follows from the fact that the SWAP gate is diagonal in the Bell basis with $\ket{\phi^{00}}$, $\ket{\phi^{01}}$, $\ket{\phi^{10}}$ being eigenvectors of eigenvalue one, and $\ket{\phi^{11}}$ being an eigenvector of eigenvalue minus one. As such, a measurement on the Bell basis plus appropriate classical post-processing allows us to compute the expectation value of the SWAP operator. 

From the previous, let us note that the measurement outcome from Fig.~\ref{fig:hadamardtest}(b) can be expressed by the tuple
$x\phi^{p_1q_1}_1 \phi^{p_2q_2}_2\ldots \phi^{p_nq_n}_n $ where $x=0,1$ denotes the measurement outcome of on the ancilla qubits, and $\phi^{p_jq_j}_j$ with $p_j,q_j=0,1$ and the Bell basis measurement outcome on the $j$-th qubits of $S_1$ and $S_2$. With this notation, $P(x\phi^{p_1q_1}_1 \phi^{p_2q_2}_2\ldots \phi^{p_nq_n}_n)$ denotes the probability of outcome $x\phi^{p_1q_1}_1 \phi^{p_2q_2}_2\ldots \phi^{p_nq_n}_n$.
If we compute the quantity
\footnotesize
\begin{equation}
   C(x)=\sum_{p_nq_n=0,1}\cdots \sum_{p_1q_1=0,1}(-1)^{\delta_{p_jq_j,11}}P(x\phi^{p_1q_1}_1 \phi^{p_2q_2}_2\ldots \phi^{p_nq_n}_n)\,,\nonumber
\end{equation}
\normalsize
for $x=0,1$, one can readily see that 
\begin{align}
    C(x)=\Tr[\dya{\psi}(\dya{x}^a\otimes {\rm SWAP}_n)]\,,
\end{align}
where ${\rm SWAP}_n$ is the SWAP operator that exchanges the state of the qubits in $S_1$ and $S_2$. An explicit evaluation leads to 
\begin{align}
    C(0)&=\frac{1}{4}( \mte{\vec{0}}{U\ad V} \mte{\vec{0}}{V\ad  U} \nonumber \\
    &+ \mte{\vec{0}}{U\ad A_{l'} A_l V} \mte{\vec{0}}{V\ad A_l\ad A_{l'}\ad U} \label{eq-HOT1}\\
    & + \Re\left[\mte{\vec{0}}{U\ad A_l V} \mte{\vec{0}}{V\ad A_{l'}\ad U}\right])\,.\nonumber
\end{align}
and
\begin{align}
    C(1)&=\frac{1}{4}( \mte{\vec{0}}{U\ad V} \mte{\vec{0}}{V\ad  U} \nonumber \\
    &+ \mte{\vec{0}}{U\ad A_{l'} A_l V} \mte{\vec{0}}{V\ad A_l\ad A_{l'}\ad U} \label{eq-HOT2}\\
    & - \Re\left[\mte{\vec{0}}{U\ad A_l V} \mte{\vec{0}}{V\ad A_{l'}\ad U}\right])\,.\nonumber
\end{align}
Then, combining \eqref{eq-HOT1} and \eqref{eq-HOT2} yields 
\begin{equation}
\Re[\gamma_{ll'}]=2(C(0)-C(1))\,.    
\end{equation}
Following a similar procedure, it can be shown that including the $R_z$ gate allows us to calculate $\Im[\gamma_{ll'}]$.

Note that the Hadamard-Overlap test can also be used to compute the real and imaginary parts of $\delta_{ll'}^{(j)}$ in \eqref{eq-deltall}. In this case an additional random unitary $R_j$ must be initially applied to the qubits in register $S_2$ in order to generate the input state $\dya{0_j} \otimes \id_{\overline{j}}$. Specifically, $R_j$ randomly applies a bit-flips to all qubits except qubit $j$:
\begin{equation}
    R_j= X_1^{r_1}\otimes X_2^{r_2}\otimes\dotsb \otimes \id_j^{r_j}\otimes\dotsb\otimes X_n^{r_n}\,,
\end{equation}
with $\vec{r}=r_1,r_2\ldots,r_n$ a random bitstring of length $n$.

\section{Proof of Proposition~\ref{Prop1}}\label{App-DQC1}

Here we prove Proposition~\ref{Prop1} of the main text, which we restate for convenience.
\begin{proposition2}
The problem of estimating the VQLS cost functions  $\CU_G$, $\CN_G$, $\CU_L$, or $\CN_L$ to within precision $\pm \delta = 1/\poly(n)$ is \DQC-hard.
\end{proposition2}

\begin{proof}
Let us first show that estimating $\CU_G$ and $\CN_G$ is \DQC-hard. Our proof is a reduction from the problem of estimating the Hilbert-Schmidt inner-product magnitude $\Delta_{\text{HS}}$ between two quantum circuits $\tilde{U}$ and $\tilde{V}$ acting on $n$-qubits \cite{QAQC}, where we have defined
\begin{align}
\Delta_{\text{HS}}(\tilde{U},\tilde{V}) :=\frac{1}{d^2}|\Tr(\tilde{V}^\dagger \tilde{U})|^2,
\end{align}
with $d=2^n$.

In particular, let us consider the following specific case of estimating $\Delta_{\text{HS}}$, which in turn can be identified as a specific instance of approximating the cost functions $\CU_G$ or $\CN_G$.  Let $A=\id$, and let $\ket{x}$ and $\ket{b}$ be $2n$-qubit states given by   
\begin{align}
\ket{x} &= V\ket{\vec{0}} = (\tilde{V} \otimes \id)E  \ket{\vec{0}}\,,\\
\ket{b} &=U\ket{\vec{0}} = (\tilde{U} \otimes \id)E  \ket{\vec{0}}\,,
\end{align}
where $E$ is an efficient unitary gate that produces a maximally entangled state (e.g., a depth-two circuit composed of Hadamard and CNOT gates). Note that here $\ket{x}$ and $\ket{b}$ correspond to the Choi states of $\tilde{V}$ and $\tilde{U}$, respectively.
The global cost function is given by
\begin{align}
 \CN_G &= \CU_G = 1 - |\ip{x}{b}|^2  \label{Eq-Fid}  \\
        &= 1-|\bra{\vec{0}}E\ad (\tilde{V}\ad\tilde{U}  \otimes \id)E \ket{\vec{0}}|^2\\ 
        &=1- \displaystyle \frac{1}{d^2}|\Tr(\tilde{V}\ad \tilde{U})|^2 = 1 - \Delta_{\text{HS}}(\tilde{U},\tilde{V})\,.
\end{align}

Moreover, it is known that approximating $\Delta_{\text{HS}}$ to within inverse polynomial precision is \DQC-hard~\cite{QAQC}, and hence the result follows. We additionally remark that estimating Eq.\  \eqref{Eq-Fid}  can also be interpreted as estimating the Fidelity between two pure states, which was also shown to be \DQC-hard \cite{cerezo2019variational}.

We now show that estimating $\CU_L$ and $\CN_L$ is also \DQC-hard. In this case our proof is reduced to the problem of approximating the trace of a unitary matrix $W$. Let $\ket{x}= V\ket{\vec{0}}$ and $\ket{b}= U\ket{\vec{0}}$ be $2n+1$-qubit states and let $A=\id$. Moreover, as depicted in Fig.~\ref{fig:DQC1}, let us denote the first qubit as $S_1$, while qubits $2,\ldots,n+1$ compose system $S_2$, and qubits $n+2,\ldots,2n+1$ compose $S_3$. Let
\begin{equation}
    U^\dagger V=  H^{S_1} C_W^{S_1 S_2} H^{S_1} E^{S_2S_3}\,,\label{eq-DQC1}
\end{equation}
where $ H^{S_1}$ is a Hadamard gate acting on $S_1$, $C_W^{S_1 S_2}$ denotes a controlled-$W$ gate with $S_1$ the control and $S_2$ the target, and $E$ is a maximally-entangling gate on $S_2$ and $S_3$.

\begin{figure}[t]
	\centering
	\includegraphics[width=.75\linewidth]{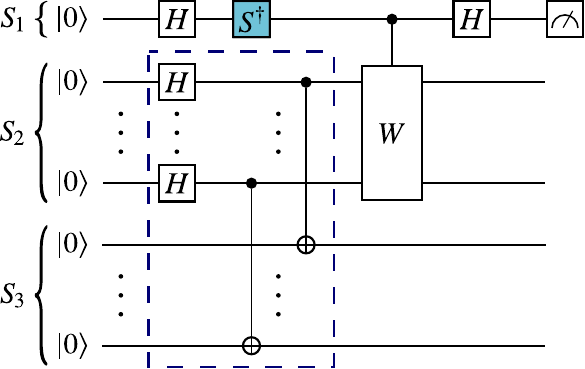}
	\caption{Schematic representation of the circuit used to compute $\CN_L$ for the specific case when $A=\id$, and $\ket{x}= V\ket{\vec{0}}$, $\ket{b}= U\ket{\vec{0}}$ are $2n+1$-qubit states such that $U^\dagger V$ is given by \eqref{eq-DQC1}. Shown is the measurement of the qubit in $S_1$. The dashed box indicates the entangling gate $E$. The phase gate in the colored box is excluded when calculating the real part of $\Tr \  W$ and included when calculating the imaginary part.}
	\label{fig:DQC1}
\end{figure}

The local cost is then
\begin{align} 
  \CN_L&=\CU_L\nonumber\\ &=  1- \frac{1}{n}\sum_{j=1}^n \mte{\vec{0}}{V\ad U \left(  \dya{0_j} \otimes \id_{\overline{j}} \right) U\ad V}\,. 
\end{align}
Consider first the case when $j=1$ and we measure the qubit in $S_1$. It is straightforward to see that the probability of measuring this qubit in the $\ket{0}$ state is $P(0)=(1+\Re(\Tr\,W))/2$.  On the other hand, the probability $P(0) = 1/2$ for all qubits in $S_2$ or $S_3$. Hence, we find 
\begin{align} 
  \CN_L&=  \frac{n+1}{2n+1}- \frac{1+\Re(\Tr\,W)}{4n+2} \,,
\end{align}
which implies that the real part of $\Tr\,W$ can be computed as
\begin{align} 
  \Re(\Tr\,W)&=  \frac{n+1}{2}- (4n+2)\CN_L -1 \,.
\end{align}
Similarly, by adding a phase gate on $S_1$ (as indicated in Fig.~\ref{fig:DQC1}) we can compute the imaginary part of $\Tr\, W$.

By choosing $U$ and $V$ according to \eqref{eq-DQC1} one finds that the problem of estimating the local cost function up to inverse
polynomial precision is equivalent to approximating the real (or imaginary) part of $\Tr\,W$. Hence, computing $\CN_L$ or $\CU_L$ is hard for \DQC, since all problems in \DQC can be seen as estimating the real part of a trace of a unitary matrix \cite{QAQC}.
\end{proof}

\section{Variable ansatz optimization}\label{ap:variable}

Here we discuss the optimization method employed for the heuristics in Section~\ref{sec:random}. As mentioned in the main text, we employed a variable-structure ansatz where the gate placement and the type of gates in $V(\vec{\alpha})$ can change during the optimization. Our approach here is similar to the variable-structure ansatzes employed in Refs.~\cite{cincio2018learning,VQSD}.

First, the gate structure and the angles of $V(\vec{\alpha})$  are randomly initialized. That is, one randomly chooses $\vec{k}$, and $\vec{\alpha}$ in~\eqref{eq-Vseq}. Then,  the optimization is performed in two alternating loops: an inner loop and an outer loop. During the inner loop, $\vec{k}$ is fixed and one optimizes over $\vec{\alpha}$. Once a local minima is reached, the circuit layout is
changed in the outer optimization loop. In this outer loop, the circuit is randomly grown by inserting into $V(\vec{\theta})$ a sequence of parametrized gates which compile to identity, such that they do not change the cost value. The previous process is then repeated by alternating between the inner and outer loops until the optimization termination condition is met. 

Here we remark that the goal of the outer loop is to enhance the expressivity  of $V(\vec{\alpha})$ and lead to smaller cost values during the next inner loop. However, it may happen that after growing the circuit, the optimizer is not able to minimize the cost function. This is due to the fact that some gate insertions do not lead to more expressive
circuits. In order to avoid such unnecessary circuit growth,  one can then accept the parametrized gate insertion conditioned to leading to smaller cost values.

\section{Gradient-based optimization}\label{ap-gradient-based}

Here we derive analytical expressions for the gradients of our cost functions, and we show that these gradients can be computed with the same circuits as those introduced in Section~\ref{Sec:cost-circuite}. This enables one to take a gradient-descent optimization approach to VQLS, and is inspired by gradient-based approaches in previous work~\cite{QAQC, mitarai2018quantum} for other VHQCAs.

For simplicity, let us first consider the global cost functions. Let us recall from \eqref{eq-Vseq} that the trainable unitary $V(\vec{\alpha})$ can be expressed as a sequence of gates $G(\alpha_i)$, where we have dropped the subscript on $G$. In turn, each gate $G(\alpha_i)$ can always be parametrized by a single-qubit rotation angle of the form $e^{-i\alpha_i \sigma_i/2}$. The gradient of $V(\vec{\alpha})$ with respect to $\vec{\alpha}$ is then given by
\begin{equation}
    \nabla_{\vec{\alpha}} V(\vec{\alpha})=\left(\frac{\partial  V(\vec{\alpha})}{\partial\alpha_1},\ldots,\frac{\partial  V(\vec{\alpha})}{\partial\alpha_D}\right)\,,
\end{equation}
and each partial derivative is
\begin{align}
    \frac{\partial  V(\vec{\alpha})}{\partial\alpha_i}&=G(\alpha_L)\ldots \frac{\partial G(\alpha_{i})}{\partial\alpha_i}\ldots G(\alpha_1)\nonumber\\
    &=-\frac{i}{2}G(\alpha_L)\ldots \sigma_i G(\alpha_{i})\ldots G(\alpha_1)\label{eq-pd-V}\,.
\end{align}

As we now show, the gradient of $\beta_{ll'}$ and $\gamma_{ll'}$ can be computed with the Hadamard-Overlap test and the Hadamard test, respectively. Let us consider the partial derivatives
\begin{align}
    \frac{\partial  \gamma_{ll'}(\vec{\alpha})}{\partial\alpha_i}&=\matl{\vec{0}}{U^\dagger A_l \frac{\partial  V(\vec{\alpha})}{\partial\alpha_i} }{\vec{0}}\matl{\vec{0}}{V(\vec{\alpha})^\dagger (A_{l'})^\dagger U}{\vec{0}}\nonumber\\
&+\matl{\vec{0}}{U^\dagger A_l V(\vec{\alpha}) }{\vec{0}}\matl{\vec{0}}{\frac{\partial  V(\vec{\alpha})^\dagger}{\partial\alpha_i} (A_{l'})^\dagger U}{\vec{0}}\,,\label{eq-pd-E}\\
\frac{\partial  \beta_{ll'}(\vec{\alpha})}{\partial\alpha_i}&=
\matl{\vec{0}}{ \frac{\partial  V(\vec{\alpha})^\dagger}{\partial\alpha_i} A_l (A_{l'})^\dagger V(\vec{\alpha}) }{\vec{0}}\nonumber\\
&+\matl{\vec{0}}{V(\vec{\alpha})^\dagger  A_l (A_{l'})^\dagger \frac{\partial  V(\vec{\alpha})}{\partial\alpha_i} }{\vec{0}}\,. \label{eq-pd-F}
\end{align}
By means of the identity $i[\sigma_i,A]=G_i(-\frac{\pi}{2})AG_i(-\frac{\pi}{2})^\dagger-G_i(\frac{\pi}{2})AG_i(\frac{\pi}{2})^\dagger$, 
which is valid for any matrix $A$, we combine Eqs.\ \eqref{eq-pd-V}--\eqref{eq-pd-F} to obtain
\begin{align}
    \frac{\partial  \gamma_{ll'}(\vec{\alpha})}{\partial\alpha_i}&=\frac{1}{2}(\matl{\vec{0}}{U^\dagger A_l   V^{i}_+(\vec{\alpha}) }{\vec{0}}\matl{\vec{0}}{V^{i}_+(\vec{\alpha})^\dagger (A_{l'})^\dagger U}{\vec{0}}\nonumber\\
&-\matl{\vec{0}}{U^\dagger A_l V^{i}_-(\vec{\alpha}) }{\vec{0}}\matl{\vec{0}}{ V^{i}_-(\vec{\alpha})^\dagger (A_{l'})^\dagger U}{\vec{0}})\,,\label{eq-pd-E2}\\
\frac{\partial  \beta_{ll'}(\vec{\alpha})}{\partial\alpha_i}&=\frac{1}{2}(
\matl{\vec{0}}{ V^{i}_+(\vec{\alpha})\ad                                                                                                                                                                                                                                                                                                                                                                                                                                     A_l (A_{l'})^\dagger V^{i}_+(\vec{\alpha}) }{\vec{0}}\nonumber\\
&-\matl{\vec{0}}{V^{i}_-(\vec{\alpha})^\dagger  A_l (A_{l'})^\dagger V^{i}_-(\vec{\alpha}) }{\vec{0}})\,, \label{eq-pd-F2}
\end{align}
where we have defined 
\begin{equation}
    V^{i}_\pm(\vec{\alpha})=G(\alpha_L)\ldots G(\alpha_{i}\pm \frac{\pi}{2})\ldots G(\alpha_1)\,.
\end{equation}
Each term in \eqref{eq-pd-E2} can be computed by means of the Hadamard-Overlap test, while the terms in \eqref{eq-pd-F2} can be determined via the Hadamard test. These results imply that the gradient with respect to $\vec{\alpha}$ of $\CN_G$ and $\CU_G$ are determined by
\begin{align}
    \frac{\partial  \CN_G}{\partial\alpha_i}&=-\frac{\sum_{l,l'm,m'} c_l c_{l'}^\ast c_m c_{m'}^\ast \frac{\partial  \gamma_{ll'}(\vec{\alpha})}{\partial\alpha_i}  \beta_{mm'}(\vec{\alpha})  }{\left(\sum_{l,l'=1}^L c_l c_{l'}^\ast    \beta_{ll'}(\vec{\alpha})\right)^2}\nonumber\\
    +&\frac{\sum_{l,l'm,m'} c_l c_{l'}^\ast c_m c_{m'}^\ast \gamma_{ll'}(\vec{\alpha})   \frac{\partial  \beta_{mm'}(\vec{\alpha})}{\partial\alpha_i}}{\left(\sum_{l,l'=1}^L c_l c_{l'}^\ast    \beta_{ll'}(\vec{\alpha})\right)^2}\,,
\end{align}
and
\begin{align}
    \frac{\partial  \CU_G}{\partial\alpha_i}&=\sum_{l,l'=1}^L c_l c_{l'}^\ast\left(\frac{\partial  \beta_{ll'}(\vec{\alpha})}{\partial\alpha_i}- \frac{\partial  \gamma_{ll'}(\vec{\alpha})}{\partial\alpha_i}\right)\,,
\end{align}
and hence that they can be computed by means of the  Hadamard-Overlap test and the Hadamard test.

A similar derivation can be used with 
\begin{align} \label{eq-deltalldp}
    \frac{\partial\delta_{ll'}^{(j)}}{\partial \alpha_i}    &= \mte{\vec{0}}{ \frac{\partial  V(\vec{\alpha})^\dagger}{\partial\alpha_i} A_{l'}\ad U (\dya{0_j} \otimes \id_{\overline{j}} ) U\ad A_l V}\nonumber\\
    &+ \mte{\vec{0}}{V\ad A_{l'}\ad U (\dya{0_j} \otimes \id_{\overline{j}} ) U\ad A_l  \frac{\partial  V(\vec{\alpha})}{\partial\alpha_i}}\,.
\end{align}
Hence the gradient of the local cost functions, $\CU_L$ and $\CN_L$, can also be computed with the circuits in Sec.~\ref{Sec:cost-circuite}.

\section{Resilience to simple noise models}\label{sec:Resilience}

Here we show that the normalized cost functions $C_G$ and $C_L$ are resilient to certain kinds of noise. In particular, we consider global depolarizing noise and measurement noise, which are relatively simple noise models. We leave the analysis of more complicated noise models for future work. 

For noise resilience, we employ the definition in Ref.~\cite{sharma2019noise}, where they define Optimal Parameter Resilience (OPR). A cost function exhibits OPR to a given noise model if this noise does not shift the global minima in parameter space, i.e., if the optimal parameters are not changed by the noise. 

OPR is very important to the success of VQLS for the following reason. If VQLS exhibits OPR, then that means that optimizing the noisy cost function can still lead to the correct optimal parameters. Then, using these correct parameters, one can prepare the state $\ket{x}$ corresponding to the correct solution of the QLSP. One can then use standard error mitigation techniques, such as zero-noise extrapolation, to obtain noise-free estimates of observable expectation values of the form $\mte{x}{O}$. Hence, OPR during the optimization process combined with error mitigation after the optimization is over can potentially lead to accurate estimation of observables on the true solution of the QLSP.

\subsection{Global Cost}

Let us write $C_G$ as:
\begin{equation}
    C_G = 1 - \frac{\nu}{\mu}
\end{equation}
where $\nu = |\ip{\psi}{b}|^2$ and $\mu = \ip{\psi}{\psi}$. For the noisy versions of these quantities, we write $ \tilde{C}_G$, $\tilde{\nu}$, and $\tilde{\mu}$. Let us first consider global depolarizing noise, which transforms the state according to $\rho \rightarrow q \rho + (1-q) \id /d$. Note that if a circuit has $L$ layers, with noise acting after each layer, then the final state is $q^L \rho + (1-q^L) \id /d$. 

First consider $\tilde{\mu}$, which is estimated by the circuit in Fig.~\ref{fig:hadamardtest}(a). We assume noise affects the estimation of the real and imaginary parts of $\tilde{\beta}_{ll'}$ to an equal extent, where $\tilde{\beta}_{ll'}$ is the noisy version of $\beta_{ll'}$. The maximally mixed state has zero expectation value for the circuit in Fig.~\ref{fig:hadamardtest}(a), which measures the Pauli $Z$ operator on the ancilla. Therefore, we obtain that $\tilde{\beta}_{ll'} = q^{L(\beta_{ll'})}\beta_{ll'}$. Here, $L(\beta_{ll'})$ is the number of layers in the circuit used to estimate $\beta_{ll'}$. Hence, we have $\tilde{\mu} = \sum_{ll'}c_l c_{l'}^* q^{L(\beta_{ll'})}\beta_{ll'}$. Finally, one can make an additional assumption that $L(\beta_{ll'}) = L_{\beta}$ is independent of $l$ and $l'$, which is a reasonable approximation since the depth of the circuit will likely be dominated by the ansatz $V$ rather than by the controlled gates in Fig.~\ref{fig:hadamardtest}(a). This gives: 
\begin{equation}\label{eqnmunoisy}
    \tilde{\mu} = q^{L_{\beta}} \mu \,.
\end{equation}

Next consider $\tilde{\nu}$, which we assume is estimated by the circuit in Fig.~\ref{fig:hadamardtest}(b). We assume noise affects the estimation of the real and imaginary parts of $\tilde{\gamma}_{ll'}$ to an equal extent, where $\tilde{\gamma}_{ll'}$ is the noisy version of $\gamma_{ll'}$. The maximally mixed state has zero expectation value for the circuit in Fig.~\ref{fig:hadamardtest}(b), which measures the Pauli $Z$ operator on the ancilla. Therefore, we obtain that $\tilde{\gamma}_{ll'} = q^{L(\gamma_{ll'})}\gamma_{ll'}$. Here, $L(\gamma_{ll'})$ is the number of layers in the circuit used to estimate $\gamma_{ll'}$. Hence, we have $\tilde{\nu} = \sum_{ll'}c_l c_{l'}^* q^{L(\gamma_{ll'})}\gamma_{ll'}$. Finally, one can make an additional assumption that $L(\gamma_{ll'}) = L_{\gamma}$ is independent of $l$ and $l'$, which is a reasonable approximation since the depth of the circuit will likely be dominated by the ansatz $V$ rather than by the controlled gates in Fig.~\ref{fig:hadamardtest}(b). This gives: 
\begin{equation}\label{eqnnunoisy}
    \tilde{\nu} = q^{L_{\gamma}} \nu \,.
\end{equation}

Combining \eqref{eqnmunoisy} and \eqref{eqnnunoisy} gives
\begin{equation}
    \tilde{C}_G = 1 - \frac{q^{L_{\gamma}}}{q^{L_{\beta}}}\frac{\nu}{\mu}\,.
\end{equation}
From this expression, we see that
\begin{equation}
    \argmin_{\vec{\alpha}}\tilde{C}_G = \argmax_{\vec{\alpha}}\left(\frac{\nu}{\mu}\right)\,.
\end{equation}
Furthermore, it is clear that
\begin{equation}
    \argmin_{\vec{\alpha}}C_G = \argmax_{\vec{\alpha}}\left(\frac{\nu}{\mu}\right)\,.
\end{equation}
Hence we arrive at
\begin{equation}
   \argmin_{\vec{\alpha}}\tilde{C}_G = \argmin_{\vec{\alpha}}C_G \,.
\end{equation}
This proves our desired statement of OPR, since it shows that the optimal parameters are unaffected by the noise.

\subsection{Local Cost}

\subsubsection{Global depolarizing noise}

Now consider the local cost $C_L$. We can write
\begin{equation}
    C_L = 1 - \frac{\omega}{\mu}\,,
\end{equation}
where $\omega = \frac{1}{n}\sum_{jll'} c_l c_{l'}^* \delta_{ll'}^{(j)}$. Let $\tilde{C}_L$, $\tilde{\omega}$, and $\tilde{\delta}_{ll'}^{(j)}$ denote the noisy versions of these quantities. 



Recall that we can expand $\delta_{ll'}^{(j)}$ according to \eqref{eq-deltall2} as $\delta_{ll'}^{(j)} = \beta_{ll'} + \zeta_{ll'}^{(j)} $, where $\zeta_{ll'}^{(j)} = \mte{\vec{0}}{V\ad A_{l'}\ad U (Z_j \otimes \id_{\overline{j}} ) U\ad A_l V}$. Also, recall that we have $\tilde{\beta}_{ll'} = q^{L_{\beta}}\beta_{ll'}$ under the same assumptions as those considered above. This gives 
\begin{align} 
\tilde{\omega} = \frac{1}{n}\sum_{jll'} c_l c_{l'}^* (q^{L_{\beta}}\beta_{ll'} + \tilde{\zeta}_{ll'}^{(j)})\,,
\end{align}
where $\tilde{\zeta}_{ll'}^{(j)}$ is the noisy version of $\zeta_{ll'}^{(j)}$.  Note that $\tilde{\zeta}_{ll'}^{(j)}$ is estimated by the circuit in Fig.~\ref{fig:hadamardtest}(c). We assume noise affects the estimation of the real and imaginary parts of $\tilde{\zeta}_{ll'}^{(j)}$ to an equal extent. The maximally mixed state has zero expectation value for the circuit in Fig.~\ref{fig:hadamardtest}(c), which measures the Pauli $Z$ operator on the ancilla. Therefore, we obtain that $\tilde{\zeta}_{ll'}^{(j)} = q^{ L( \zeta_{ll'}^{(j)} ) } \zeta_{ll'}^{(j)}$. Here, $ L( \zeta_{ll'}^{(j)} ) $ is the number of layers in the circuit used to estimate $\zeta_{ll'}^{(j)}$. Finally, one can make an additional assumption that $L( \zeta_{ll'}^{(j)} )  = L_{\zeta}$ is independent of $l$, $l'$, and $j$, which is a reasonable approximation since the depth of the circuit will likely be dominated by either the ansatz $V$ or the unitary $U$, rather than by the controlled gates in Fig.~\ref{fig:hadamardtest}(c). This gives $\tilde{\zeta}_{ll'}^{(j)} = q^{ L_{\zeta} } \zeta_{ll'}^{(j)}$, which implies
\begin{align} 
\tilde{\omega} &= q^{L_{\beta}} \frac{1}{n}\sum_{jll'} c_l c_{l'}^* (\beta_{ll'} +  \frac{q^{ L_{\zeta} }}{q^{L_{\beta}}}\zeta_{ll'}^{(j)})\notag \\
&= q^{L_{\beta}} \frac{1}{n}\sum_{j} \Tr(\dya{\chi} (\rho_j \ot \id_{\overline{j}}  ) )\,,
\end{align}
where $\ket{\chi} = U\ad A\ket{x}$ is an unnormalized state, and $\rho_j = \frac{1}{2}(\id_j + \frac{ q^{ L_{\zeta}} }{q^{L_{\beta}} } Z_j)$ is a Hermitian (but not necessarily positive semidefinite) matrix. Note that we have $\tilde{\mu} = q^{L_{\beta}} \mu $, which gives:
\begin{align} 
\tilde{C}_L &= 1 - \frac{1}{\mu}\frac{1}{n}\sum_j \Tr(\dya{\chi} (\rho_j \ot \id_{\overline{j}}  ) )\notag \\
&= 1 - \frac{1}{n}\sum_j \Tr(\dya{\hat{\chi}} (\rho_j \ot \id_{\overline{j}}  ) )\,,
\end{align}
where $\ket{\hat{\chi}} = \ket{\chi}/\sqrt{\mu}$ is a normalized state.

Let us now argue that $\tilde{C}_L$ is a faithful cost function, reaching its minimum value iff $A\ket{x}\propto \ket{b}$. Let us first note that the minimum value of the cost can be found as follows. Note that $\Tr(\dya{\chi} (\rho_j \ot \id_{\overline{j}}  ))$ is not larger than the largest eigenvalue of $(\rho_j \ot \id_{\overline{j}}  )$, and this eigenvalue is $(1+q^{L_{\zeta}} / q^{L_{\beta}}  )/2$. Hence, we have 
\begin{align} 
\tilde{C}_L \geq (1-q^{L_{\zeta}} / q^{L_{\beta}}  )/2\,.
\end{align}
The lower bound here is actually achievable and hence corresponds to the minimum value of $\tilde{C}_L$. Let us denote this minimum value as $\tilde{C}_L^{\min}$.  

Namely, if we assume that $\ket{x}$ is a solution to the QLSP, then we have $A\ket{x}\propto \ket{b}$, which implies $\ket{\chi}\propto \ket{\vec{0}}$ and $\ket{\hat{\chi}} = \ket{\vec{0}}$. In turn, this implies $\Tr(\dya{\chi} (\rho_j \ot \id_{\overline{j}}  )) = (1+q^{L_{\zeta}} / q^{L_{\beta}}  )/2$ and hence
$\tilde{C}_L = \tilde{C}_L^{\min}$.

Conversely, if we assume that $\tilde{C}_L = \tilde{C}_L^{\min}$, then this implies that $\ket{\hat{\chi}}$ must be an eigenvector of $(\rho_j \ot \id_{\overline{j}}  )$ for all $j$, with an eigenvalue of $(1+q^{L_{\zeta}} / q^{L_{\beta}}  )/2$. This implies that $\Tr_{\overline{j}}(\dya{\hat{\chi}}) = \dya{0}$ for all $j$, i.e., that $\ket{\hat{\chi}}$ is locally the zero state on each qubit. The only state that satisfies this is $\ket{\hat{\chi}} = \ket{\vec{0}}$.  This state corresponds to $A \ket{x} \propto \ket{b}$, and hence the QLSP is solved in this case. 

This completes the argument that $\tilde{C}_L$ is a faithful cost function. Note that this faithfulness property corresponds to OPR, assuming the ansatz $V(\vec{\alpha})$ is capable of expressing the true solution. Hence we have proven OPR for $\tilde{C}_L$ under global depolarizing noise.

\subsubsection{Measurement noise}

Finally, we remark that the above proof has implications for measurement noise as well. Measurement noise that is symmetric with respect to the 0 and 1 outcomes can be modeled as a local depolarizing channel acting on the state just prior to the measurement. Moreover, if only a single qubit is being measured, then this local depolarizing channel can be replaced by a global depolarizing channel. We remark that this is the case for the circuits in Fig.~\ref{fig:hadamardtest}(a) and (c), which are the circuits used to compute the local cost $\tilde{C}_L$. Hence, for this local cost, any symmetric measurement noise can mathematically be viewed as an additional global depolarizing channel acting just prior to measurement. This means that $\tilde{C}_L$ exhibits OPR to a noise model that includes both global depolarizing as well as symmetric measurement noise.

\section*{Acknowledgements}

We thank Rolando Somma for helpful conversations. We thank Rigetti for providing access to their quantum computer. The views expressed in this article are those of the authors and do not reflect those of Rigetti. CBP acknowledges support from the U.S. Department of Energy (DOE) through a quantum computing program sponsored by the Los Alamos National Laboratory (LANL) Information Science \& Technology Institute. MC was supported by the Center for Nonlinear Studies at LANL. YS and PJC acknowledge support from the LANL ASC Beyond Moore's Law project. MC, YS, LC, and PJC also acknowledge support from the LDRD program at LANL. LC was supported by the DOE through the J. Robert Oppenheimer fellowship. This work was also supported by the U.S. DOE, Office of Science, Office of Advanced Scientific Computing Research, under the Quantum Computing Application Teams program.

\medskip



%


\clearpage

\newpage

\tikzstyle{decision} = [diamond, draw, fill=purple!60, 
    text width=4.5em, text badly centered, node distance=3cm, inner sep=0pt, rounded corners, minimum height=7em, minimum width = 7em]
\tikzstyle{block} = [rectangle, draw, fill=blue!60, 
    text width=5em, text centered, rounded corners, minimum height=4em]
\tikzstyle{block2} = [rectangle, draw, fill=orange!60, 
    text width=5em, text centered, rounded corners, minimum height=4em]
\tikzstyle{line} = [draw, -latex']
\tikzstyle{cloud} = [draw, ellipse, fill=pink!150, node distance=3cm,
    minimum height=2.5em]

\onecolumngrid

\appendix

\begin{center}
\Large{Supplemental Material for \\ ``Variational Quantum Linear Solver''}
\end{center}

\twocolumngrid


\renewcommand{\figurename}{Supplementary Figure}
\renewcommand{\tablename}{Supplementary Table}
\setcounter{figure}{0}
\renewcommand{\thefigure}{S\arabic{figure}}
\renewcommand{\theHfigure}{S\arabic{figure}}
\renewcommand{\thetable}{S\arabic{table}}
\renewcommand{\theequation}{S\arabic{equation}}

\section{Supplementary Note 1 -- Additional Implementations}

\subsection{Scaling Heuristics for $8\times 8$ Systems}

\begin{figure}[t]
	\centering
	\includegraphics[width=.8\columnwidth]{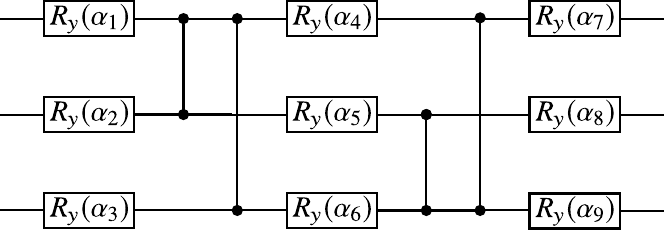}
	\caption{Hardware-Efficient Ansatz used to solve the QLSPs in~\eqref{eqn:A1}--\eqref{eqn:Examplebstate}. Since $A$ and $\ket{b}$ are real, $V(\vec{\alpha})$ contains only rotation around the $y$-axis $R_y (\alpha_i)= e^{-i\alpha_i Y/2}$, and control-$Z$ gates. \label{fig:ansatz8x8}}
\end{figure}

Here we study the scaling of the VQLS algorithm for $8\times 8$ systems. For this purpose we  employed the ansatz of Fig.~\ref{fig:ansatz8x8} (with randomly initialized parameters) and we numerically implement VQLS to solve the three different QLSPs with different degeneracy $g$ in the minimum eigenvalue of $A$ defined by:
\begin{align}
    A_1 &= \frac{1}{8 \kappa} [ 4 (\kappa + 1) \id + (\kappa -1)(  Z_3 + Z_2 + 2 Z_1) ]
\label{eqn:A1} \\
    A_2 &= \frac{1}{4 \kappa} \left[ 2 (\kappa + 1) \id + (\kappa - 1) (Z_3  + Z_2) \right]
\label{eqn:A2} \\
    A_3 &= \frac{1}{2 \kappa} \left[ (\kappa + 1) \id + (\kappa - 1) Z_3 \right] \,.
\label{eqn:A3} 
\end{align}
The degeneracy of each matrix is $g=1,2,4$, respectively. We remark that we considered different values of $g$ to analyze if this parameter affects the VQLS performance. 
The state $\ket{b}$~is
\begin{equation}
\label{eqn:Examplebstate} 
    \ket{b} = H^{\otimes 3} \ket{000}\,.
\end{equation}

\begin{figure*}[t]
	\centering
	\includegraphics[width=\textwidth]{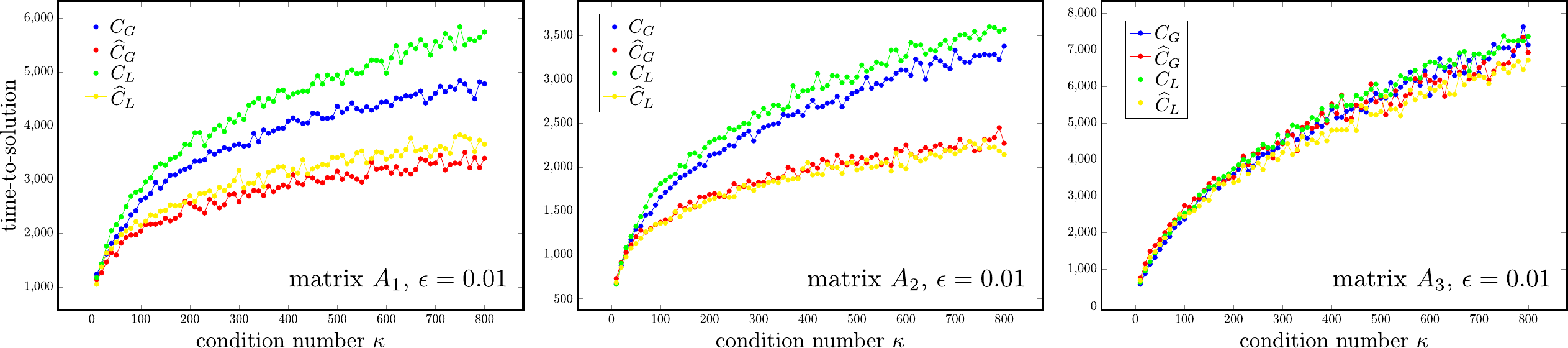}
	\caption{Time-to-solution versus condition number $\kappa$. The time-to-solution is the mean number of executions needed to guarantee a desired precision $\epsilon$. The QLSP is determined by $\ket{b} $ of \eqref{eqn:Examplebstate}, and $A$ given by: 
    {\sl Left:} Matrix $A_1$ of~\eqref{eqn:A1}. {\sl Center:} Matrix $A_2$ of~\eqref{eqn:A2}. {\sl Right:} Matrix $A_3$ of~\eqref{eqn:A3}. For each data point we ran and averaged 1000 instances of the VQLS algorithm. In all cases we trained the gate sequence by minimizing  $\CU_G$, $\CN_G$, $\CU_L$, and $\CN_L$. As can be seen, the scaling in terms of the condition number $\kappa$ appears to be efficient for all $A$ matrices and for all cost functions. \label{fig:gvskappa8x8}}
\end{figure*}

Figure~\ref{fig:gvskappa8x8} shows results  plotting time-to-solution versus $\kappa$ for the aforementioned $A$ matrices by training $\CU_G$, $\CN_G$, $\CU_L$, and $\CN_L$.  In all cases we employed the operational meanings of our cost functions in~\eqref{eq-costfunctionsbounds} and \eqref{eq-costfunctionsboundsTightened} of the main text for our certification procedure, i.e., to upper-bound the quantity $\epsilon$. For each data point in Fig.~\ref{fig:gvskappa8x8}, we implemented and averaged over 1000 runs of VQLS.

These results show that the $\kappa$ scaling is efficient for the problems considered  (regardless of the value of $g$ considered). This is  in agreement with the scaling observed in the main text. It is worth noting that this efficient scaling holds for all of our cost functions. The unnormalized cost functions have slightly better performance for the $A_1$ and $A_2$ matrices, although all four cost functions perform similarly for $A_3$, indicating that the performances of different cost functions is problem dependent.

\subsection{Implementation with QAOA Ansatz}

\begin{figure}[t]
    \centering
    \includegraphics[width=.8\columnwidth]{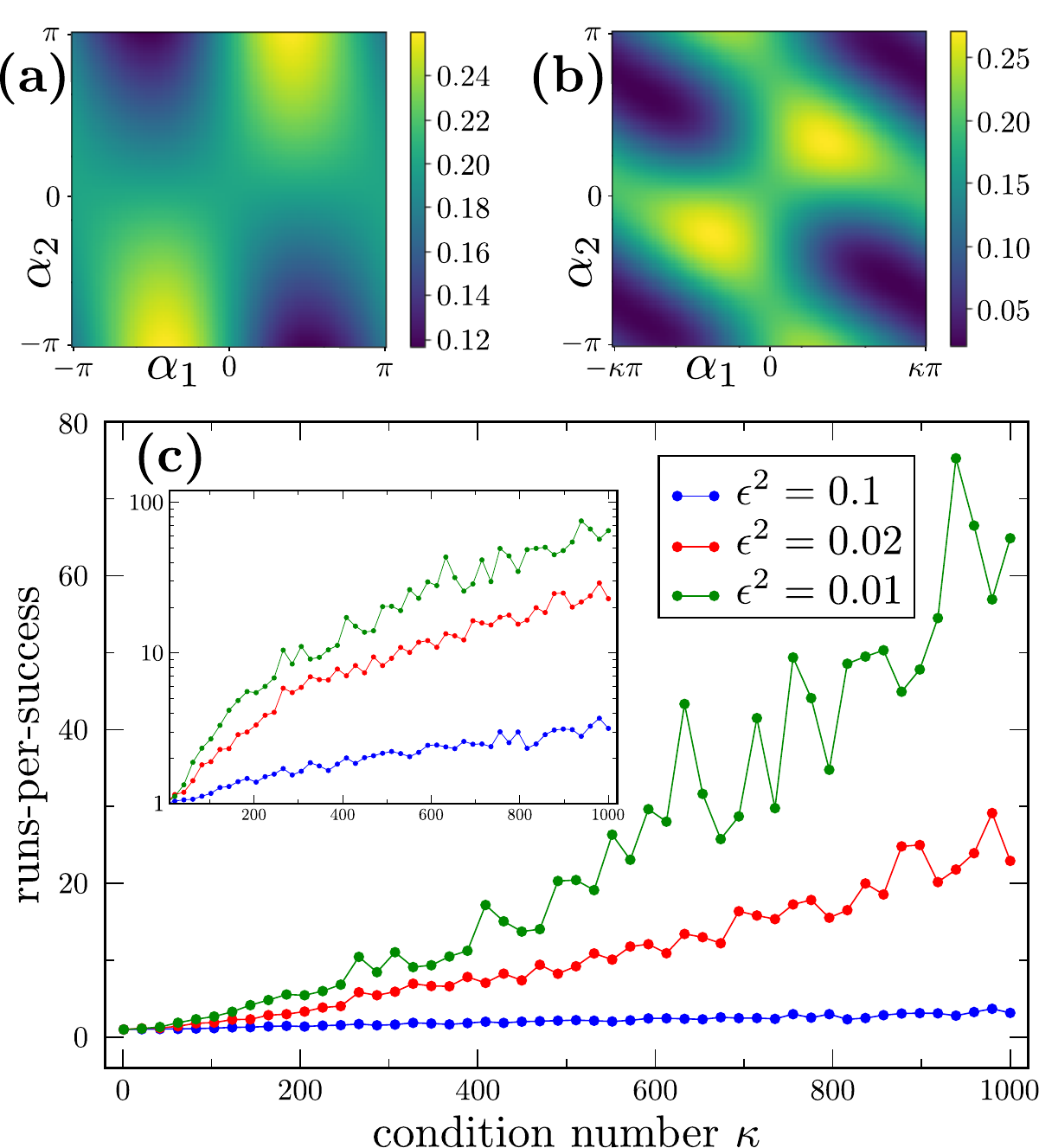}
    \caption{(a) Landscape for $\CU_G$ with a QAOA ansatz of $p = 1$ layer and unscaled parameters $\alpha_1$ and $\alpha_2$. Here, $\alpha_1$ ($\alpha_2$) corresponds the the parameter in the driver (mixer) Hamiltonian. (b) Landscape for $\CU_G$ with a QAOA ansatz of $p = 1$ layer where $\alpha_1$ was scaled by the condition number $\kappa$. In both cases the QLSP is defined by a randomly generated $4 \times 4$ matrix with condition number $\kappa \approx 11$, and with $\ket{b}$ given by \eqref{eqn:Examplebstate}. 
    The scaled landscape contains more regions of low cost and thus makes optimization more likely to be successful.  (c)~ Time-to-solution versus condition number $\kappa$ for the QLSP on three qubits defined by the  $A_2$ matrix of~\eqref{eqn:A2} and with $\ket{b}$ given by~\eqref{eqn:Examplebstate}. Three curves are shown for $\epsilon^2 = 0.10, 0.02$, and $0.01$. The inset depicts the same data on a logarithmic scale. As can be seen from the inset, the scaling in $\kappa$ is sub-exponential for each $\epsilon$ considered.}   
    \label{fig:qaoa-cost-landscape}
\end{figure}

Here we numerically analyze the VQLS scaling with $\kappa$ when employing the QAOA ansatz. Since poorly conditioned matrices (i.e., large $\kappa$) are more difficult to invert, we expect that for fixed $\epsilon$ the number of layers $p$  must increase with $\kappa$.  While this is generally true, we can also alleviate this issue by evolving with the driver Hamiltonian $H_D$ for a longer time. This corresponds to scaling the parameters $\alpha_i $ for odd $i$ in \eqref{eqn:qaoa-ansatz} of the main text by some value that grows with $\kappa$. As shown in Fig.~\ref{fig:qaoa-cost-landscape}(a) and (b), this scaling can indeed transform the cost landscape such that it contains more regions of low cost and thus makes optimization more likely to be successful. 

In Fig.~\ref{fig:qaoa-cost-landscape}(c), we show the  time-to-solution versus the condition number. Here, we consider the QLSP on three qubits defined by the  $A_2$ matrix of~\eqref{eqn:A2} and with $\ket{b}$ given by~\eqref{eqn:Examplebstate}. For this small scale-implementation we obtained the time-to-solution be exactly computing $\epsilon$. The condition number was varied from $\kappa = 10^{0}$ to $\kappa = 10^{3}$. For each $\kappa$, VQLS was implemented 100 times with the parameters randomly initialized.  
For each of the three values of $\epsilon$ considered, the scaling with $\kappa$ is sub-exponential. Hence, these results indicate that VQLS with QAOA also scales efficiently in the condition number $\kappa$. Finally, we emphasize that these results were obtained with only $p = 1$ round of QAOA, and remark that additional rounds $p > 1$ may lead to better performance.

\subsection{Implementations on Rigetti's quantum computer}
\label{Implementations}

Here we present additional implementations performed on Rigetti's quantum device \textit{16Q Aspen-4}. We have considered different problem sizes, from $2\times2$ up to $32\times32$. We additionally recall that the matrices $A$ and states $\ket{b}$ in these QLSP are such that the ansatz and the cost computing circuits are simplified. 

First we present the results of a $32 \times 32$  (i.e., 5-qubit) implementation of VQLS using Rigetti's quantum chip \textit{16Q Aspen-4}. We considered the QLSP defined by
\begin{equation}\label{eq-A-Rigetti}
    A = \id + 0.2 X_1Z_2 +0.2X_1\,,
\end{equation}
and $\ket{b} = H_1 H_3 H_4 H_5 \ket{0}^{\otimes 5}$. This particular choice of $A$ and $\ket{b}$ is motivated from the fact that they lead to simplified ansatz and cost evaluation circuits. In particular, the ansatz considered consists of $R_y(\alpha_i)$ gates acting on each qubit.

\begin{figure}[t]
	\centering
	\includegraphics[width=.9\columnwidth]{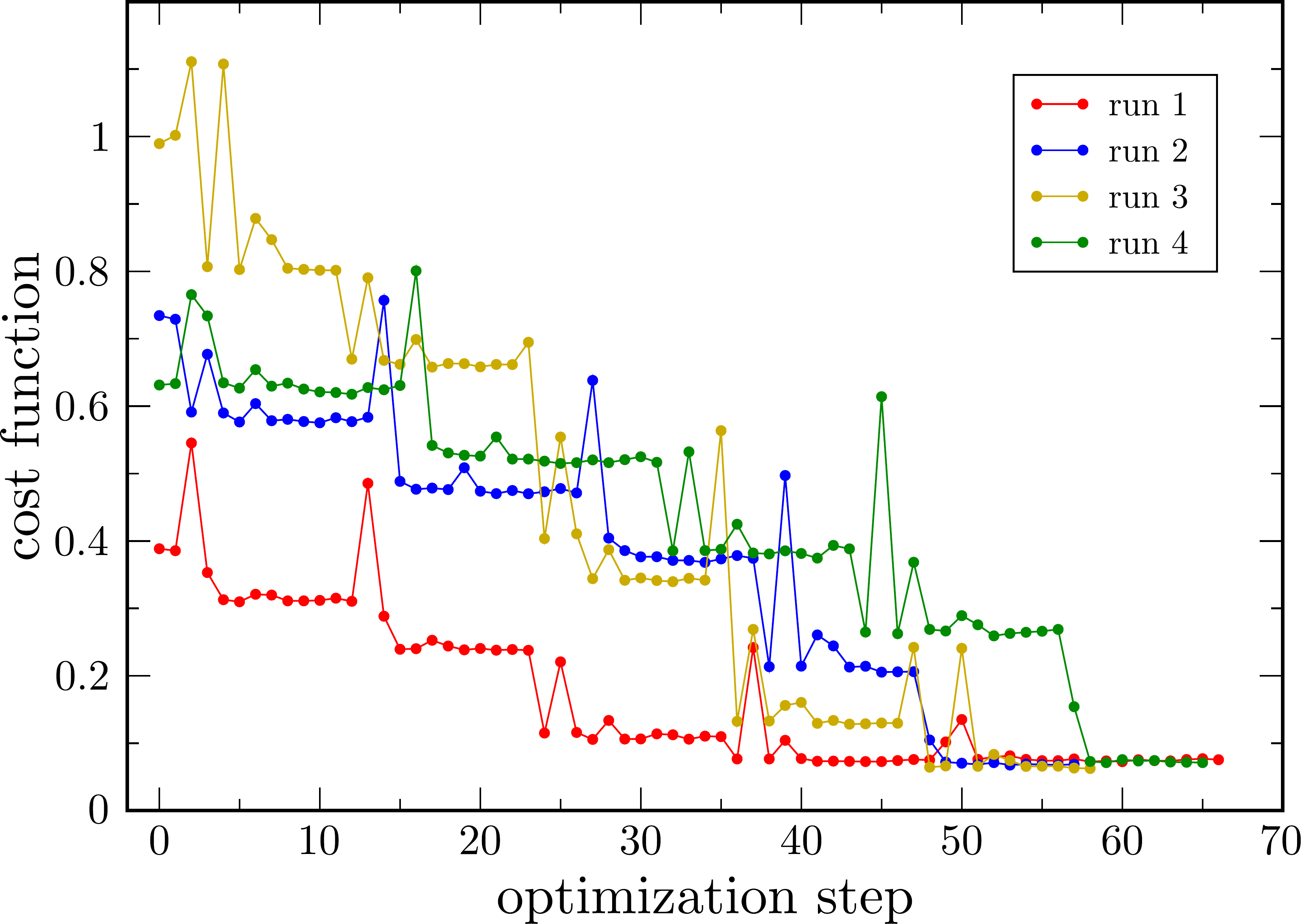}
	\caption{Implementation of VQLS on Rigetti's quantum hardware. Cost function $\CN_L$ is plotted versus number of optimization steps, where $A$ is given by \eqref{eq-A-Rigetti}.  One can observe that for every run the cost function is reduced to a value of~$\sim~7~\times~10^{-2}$. Due to noise present in the quantum device the cost does not go to zero. \label{fig:5q3terms}}
\end{figure}

The results are shown in Fig.~\ref{fig:5q3terms}. At each run of the VQLS algorithm the parameters were initialized to random angles, and the classical optimization was performed with the Powell method \cite{powell1978fast}. For every run, the local cost function $\CN_L$ of~\eqref{eqn:local-cost1} achieved a value of $\sim 7\times10^{-2}$ (hardware noise prevented further cost reduction). While this cost value led to a trivial bound on $\epsilon$ via \eqref{eq-costfunctionsbounds}, we nevertheless found the solution $\ket{x}$ to be of high quality. We verified this by measuring the expectation value of different Hermitian observables $M$ on the state $\ket{x}$ prepared on the quantum computer. According to~\eqref{eqn:DMdefinition}, we can use $D(M)^2$ as a figure of merit to quantify the quality of our solution. For all $M$ we considered, $D(M)^2$ was no larger than 0.01, and hence the results have a good agreement with the exact solution.  See Table~\ref{fig:5qtable3terms} for all values of $D(M)^2$.

\begin{figure}[h!]
\hspace*{0.6cm}
	\centering
 \captionsetup{margin={0.0cm,-8.7cm}}
	\includegraphics[width=0.9\textwidth]{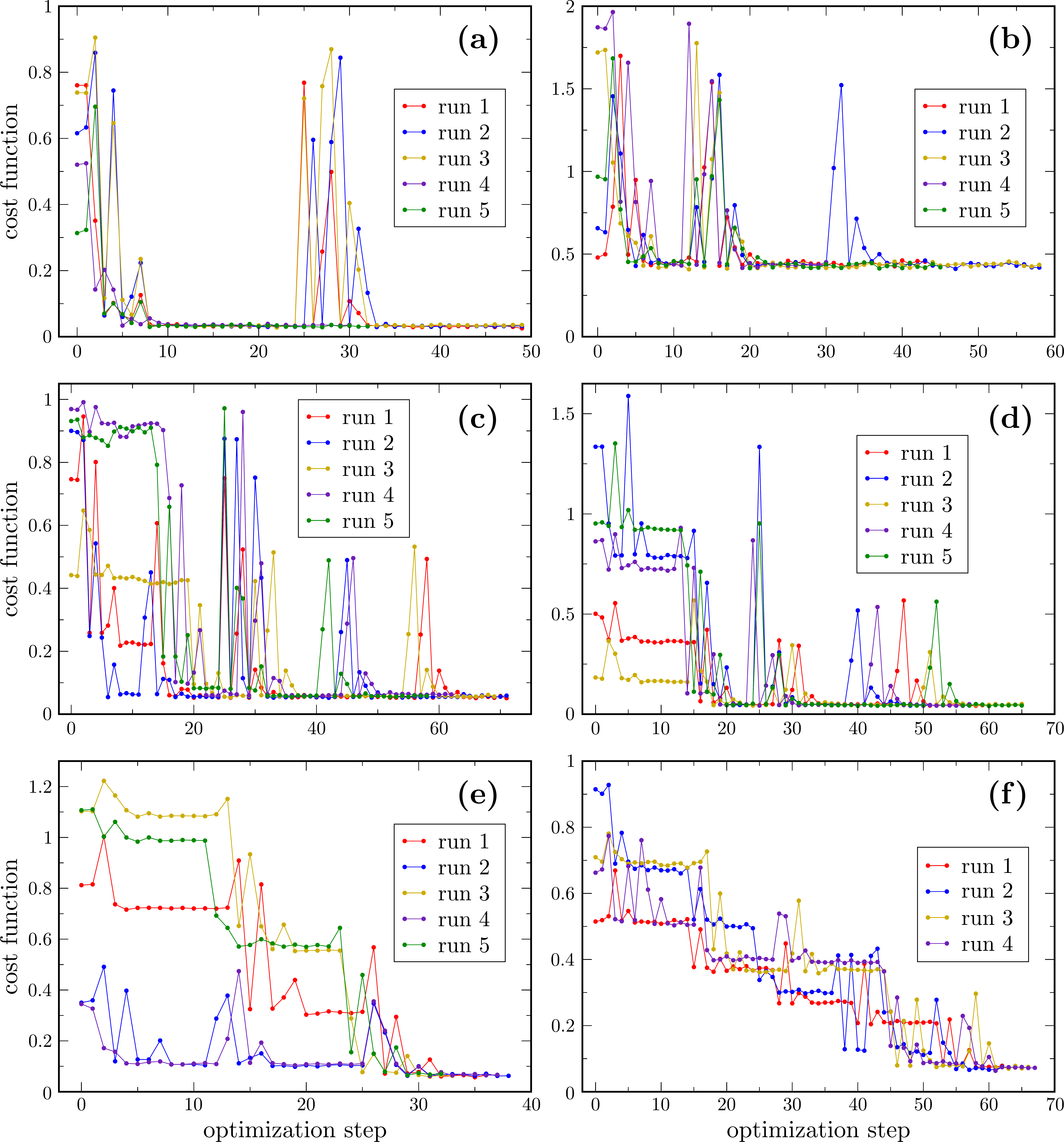}
	\caption{\hspace{0.6cm}Cost function versus number of optimization steps. The classical optimization algorithm employed is the Powell method which uses an unconstrained bi-directional search. Randomization in this algorithm occasionally leads to spikes in the cost function, visible in the plots, which quickly deteriorate as the optimizer reverts back towards better parameters. (a) $A_{2\times2} = H$, and $\ket{b} = X \ket{0}$. (b) $A_{2\times2} = \id + 0.25 Z$, and $\ket{b} = X \ket{0}$. (c) $A_{4\times4} = X_1 H_2$, and $\ket{b} = H_1 H_2 \ket{\vec{0}}$. (d) $A_{4\times4} = \id + 0.25 Z_2$, and $\ket{b} = H_1 \ket{\vec{0}}$. (e) $A_{8\times8} = \id + 0.25 Z_3$, and $\ket{b} = H_1 H_2 \ket{\vec{0}}$. (f) $A_{32\times32} = \id + 0.25 X_5$, and $\ket{b} = H^{\otimes 5}\ket{\vec{0}}$. \label{fig:cost}}
\end{figure}

Figure~\ref{fig:cost} shows the value of the cost function versus the number of optimization steps for different linear systems and for several runs. It is worth mentioning that the cost function is reduced to values $\lesssim 0.1$ for every example, except for the case depicted in panel (b). In this particular case, the solution of the $2\times2$ linear system is $\ket{x_0} = \ket{1}$. Therefore, one may note the effect of relaxation to the state $\ket{0}$ in the quantum device, which likely significantly affected the result quality. The  Tables ~\ref{fig:1qtable},~\ref{fig:1qtable2},~\ref{fig:1qtable3},~\ref{fig:1qtable4},~\ref{fig:1qtable5} and~\ref{fig:5qtable} correspond to the examples shown in Figure~\ref{fig:cost}. In the tables we show the expectation values of several observables $M$, obtained from the output of the VQLS and we compare them to the exact ones.

\begin{table}[t]
\begin{tabular}{|c|c|c|c|}
\hline
$M$ & $\ave{M}_{\text{exact}}$ & $\ave{M}_{\text{exp}}$ & $D(M)^2$ \\ \hline
$Z$ & 0 & 0.04 $\pm$ 0.02 & 0.002 $\pm$ 0.002 \\ \hline
\end{tabular}
\caption{Expectation value of an observable $M$ computed with the exact solution, and with the output solution of  VQLS. $D(M)$ measures the difference between these two results. The linear system considered is  $A_{2\times2} = H$, and $\ket{b} = X \ket{0} . $\label{fig:1qtable}}
\end{table}

\begin{table}[t]
\begin{tabular}{|c|c|c|c|}
\hline
$M$ & $\ave{M}_{\text{exact}}$ & $\ave{M}_{\text{exp}}$ & $D(M)^2$ \\ \hline
$Z$ & -1 & -0.819 $\pm$ 0.005 & 0.032 $\pm$ 0.002 \\ \hline
\end{tabular}
\caption{Expectation value of an observable $M$ computed with the exact solution, and with the output solution of  VQLS. $D(M)$ measures the difference between these two results. The linear system considered is  $A_{2\times2} = \id$ + 0.25 $Z$, and $\ket{b} = X \ket{0}$. \label{fig:1qtable2}}
\end{table}

\begin{table}[t]
\begin{tabular}{|c|c|c|c|}
\hline
$M$ & $\ave{M}_{\text{exact}}$ & $\ave{M}_{\text{exp}}$ & $D(M)^2$ \\ \hline
$Z_2$ & 1 & 0.943 $\pm$ 0.003 & 0.0032 $\pm$ 0.0003 \\ \hline
$Z_1$ & 0 & 0.02 $\pm$ 0.04 & 0.000 $\pm$ 0.002 \\ \hline
$Z_1Z_2$ & 0 & 0.02 $\pm$ 0.04 & 0.000 $\pm$ 0.002 \\ \hline
\end{tabular}
\caption{Expectation value of observables $M$ computed with the exact solution, and with the output solution of  VQLS. $D(M)$ measures the difference between these two results. The linear system considered is  $A_{4\times4} = X_1 H_2$, and $\ket{b} = H_1 H_2 \ket{\vec{0}}$. \label{fig:1qtable3}}
\end{table}

\begin{table}[t]
\begin{tabular}{|c|c|c|c|}
\hline
$M$ & $\ave{M}_{\text{exact}}$ & $\ave{M}_{\text{exp}}$ & $D(M)^2$ \\ \hline
$Z_2$ & 1 & 0.930 $\pm$ 0.004 & 0.0047 $\pm$ 0.0005 \\ \hline
$Z_1$ & 0 & 0.00 $\pm$ 0.02 & 0.00000 $\pm$ 0.00006 \\ \hline
$Z_1Z_2$ & 0 & 0.00 $\pm$ 0.02 & 0.00000 $\pm$ 0.00009 \\ \hline
\end{tabular}
\caption{Expectation value of observables $M$ computed with the exact solution, and with the output solution of  VQLS. $D(M)$ measures the difference between these two results. The linear system considered is  $A_{4\times4} = \id + 0.25 Z_2$, and $\ket{b} = H_1 \ket{\vec{0}}$. \label{fig:1qtable4}}
\end{table}

\begin{table}[t]
\begin{tabular}{|c|c|c|c|}
\hline
$M$ & $\ave{M}_{\text{exact}}$ & $\ave{M}_{\text{exp}}$ & $D(M)^2$ \\ \hline
$Z_3$ & 1 & 0.88 $\pm$ 0.01 & 0.013 $\pm$ 0.002 \\ \hline
$Z_2$ & 0 & 0.00 $\pm$ 0.04 & 0.0000 $\pm$ 0.0001 \\ \hline
$Z_2Z_3$ & 0 & 0.00 $\pm$ 0.03 & 0.0000 $\pm$ 0.0002 \\ \hline
$Z_1$ & 0 & 0.04 $\pm$ 0.05 & 0.002 $\pm$ 0.003 \\ \hline
$Z_1Z_3$ & 0 & 0.04 $\pm$ 0.04 & 0.002 $\pm$ 0.004 \\ \hline
$Z_1Z_2$ & 0 & -0.002 $\pm$ 0.009 & 0.00000 $\pm$ 0.00004 \\ \hline
$Z_1Z_2Z_3$ & 0 & -0.004 $\pm$ 0.009 & 0.00002 $\pm$ 0.00009 \\ \hline
\end{tabular}
\caption{Expectation value of observables $M$ computed with the exact solution, and with the output solution of  VQLS. $D(M)$ measures the difference between these two results. The linear system considered is  $A_{8\times8} = \id + 0.25Z_3$, and $\ket{b} = H_1 H_2 \ket{\vec{0}}$. \label{fig:1qtable5}}
\end{table}

\begin{table}[t]
\footnotesize{
\begin{tabular}{|c|c|c|c|}
\hline
$M$ & $\ave{M}_{\text{exact}}$ & $\ave{M}_{\text{exp}}$ & $D(M)^2$ \\ \hline
$Z_5$ & 0 & 0.180 $\pm$ 0.030    & 0.030000 $\pm$ 0.01000 \\ \hline
$Z_4$ & 0 & 0.000 $\pm$ 0.100    & 0.000000 $\pm$ 0.00400 \\ \hline
$Z_4Z_5$ & 0 & 0.000 $\pm$ 0.020    & 0.000000 $\pm$ 0.00040 \\ \hline
$Z_3$ & 0 & 0.000 $\pm$ 0.100    & 0.000000 $\pm$ 0.00900 \\ \hline
$Z_3Z_5$ & 0 & 0.000 $\pm$ 0.020    & 0.000000 $\pm$ 0.00040 \\ \hline
$Z_3Z_4$ & 0 & -0.006 $\pm$ 0.009   & 0.000000 $\pm$ 0.00010 \\ \hline
$Z_3Z_4Z_5$ & 0 & 0.000 $\pm$ 0.001    & 0.000000 $\pm$ 0.000001 \\ \hline
$Z_2$ & 0 & 0.100 $\pm$ 0.020    & 0.010000 $\pm$ 0.00500 \\ \hline
$Z_2Z_5$ & 0 & 0.019 $\pm$ 0.009    & 0.000300 $\pm$ 0.00030 \\ \hline
$Z_2Z_4$ & 0 & 0.000 $\pm$ 0.010    & 0.000000 $\pm$ 0.00004 \\ \hline
$Z_2Z_4Z_5$ & 0 & -0.001 $\pm$ 0.009   & 0.000000 $\pm$ 0.00003 \\ \hline
$Z_2Z_3$ & 0 & 0.000 $\pm$ 0.010    & 0.000000 $\pm$ 0.00002 \\ \hline
$Z_2Z_3Z_5$ & 0 & 0.005 $\pm$ 0.004    & 0.000020 $\pm$ 0.00004 \\ \hline
$Z_2Z_3Z_4$ & 0 & -0.007 $\pm$ 0.006   & 0.000050 $\pm$ 0.00009 \\ \hline
$Z_2Z_3Z_4Z_5$ & 0 & -0.002 $\pm$ 0.008   & 0.000000 $\pm$ 0.00004 \\ \hline
$Z_1$ & 0 & 0.010 $\pm$ 0.020    & 0.000200 $\pm$ 0.00070 \\ \hline
$Z_1Z_5$ & 0 & 0.005 $\pm$ 0.008    & 0.000030 $\pm$ 0.00009 \\ \hline
$Z_1Z_4$ & 0 & 0.003 $\pm$ 0.005    & 0.000010 $\pm$ 0.00003 \\ \hline
$Z_1Z_4Z_5$ & 0 & -0.002 $\pm$ 0.007   & 0.000000 $\pm$ 0.00004 \\ \hline
$Z_1Z_3$ & 0 & 0.000 $\pm$ 0.010    & 0.000000 $\pm$ 0.00001 \\ \hline
$Z_1Z_3Z_5$ & 0 & -0.002 $\pm$ 0.005   & 0.000000 $\pm$ 0.00002 \\ \hline
$Z_1Z_3Z_4$ & 0 & 0.000 $\pm$ 0.008    & 0.000000 $\pm$ 0.00001 \\ \hline
$Z_1Z_3Z_4Z_5$ & 0 & -0.001 $\pm$ 0.004   & 0.000000 $\pm$ 0.00004 \\ \hline
$Z_1Z_2$ & 0 & 0.006 $\pm$ 0.004    & 0.000040 $\pm$ 0.00005 \\ \hline
$Z_1Z_2Z_5$ & 0 & 0.000 $\pm$ 0.010    & 0.000010 $\pm$ 0.00007 \\ \hline
$Z_1Z_2Z_4$ & 0 & -0.002 $\pm$ 0.001   & 0.000000 $\pm$ 0.00001 \\ \hline
$Z_1Z_2Z_4Z_5$ & 0 & 0.002 $\pm$ 0.005    & 0.000000 $\pm$ 0.00002 \\ \hline
$Z_1Z_2Z_3$ & 0 & 0.001 $\pm$ 0.007    & 0.000000 $\pm$ 0.00002 \\ \hline
$Z_1Z_2Z_3Z_5$ & 0 & 0.000 $\pm$ 0.006    & 0.000000 $\pm$ 0.00001 \\ \hline
$Z_1Z_2Z_3Z_4$ & 0 & 0.001 $\pm$ 0.002    & 0.000001 $\pm$ 0.000006 \\ \hline
$Z_1Z_2Z_3Z_4Z_5$ & 0 & 0.002 $\pm$ 0.001    & 0.000004 $\pm$ 0.000007 \\ \hline
\end{tabular}
}
\caption{Expectation value of observables $M$ computed with the exact solution, and with the output solution of  VQLS. $D(M)$ measures the difference between these two results. The linear system considered is $A_{32\times32} = \id + 0.25 X_5$, and $\ket{b} = H^{\otimes 5}\ket{\vec{0}}$. \label{fig:5qtable}}
\end{table}

\begin{table}[t]
\footnotesize{
\begin{tabular}{|c|c|c|c|}
\hline
$M$ & $\ave{M}_{\text{exact}}$ & $\ave{M}_{\text{exp}}$ & $D(M)^2$ \\ \hline
$Z_5$ & 0 & 0.1 $\pm$ 0.1 & 0.01 $\pm$ 0.04 \\ \hline
$Z_4$ & 0 & 0.00 $\pm$ 0.04 & 0.0 $\pm$ 0.0005 \\ \hline
$Z_4Z_5$ & 0 & 0.00 $\pm$ 0.01 & 0.0 $\pm$ 0.00007 \\ \hline
$Z_3$ & 0 & 0.0 $\pm$ 0.1 & 0.0 $\pm$ 0.004 \\ \hline
$Z_3Z_5$ & 0 & 0.01 $\pm$ 0.02 & 0.0002 $\pm$ 0.0009 \\ \hline
$Z_3Z_4$ & 0 & -0.002 $\pm$ 0.007 & 0.0 $\pm$ 0.00003 \\ \hline
$Z_3Z_4Z_5$ & 0 & 0.000 $\pm$ 0.005 & 0.0 $\pm$ 0.000009 \\ \hline
$Z_2$ & 1 & 0.971 $\pm$ 0.002 & 0.0008 $\pm$ 0.0001 \\ \hline
$Z_2Z_5$ & 0 & 0.1 $\pm$ 0.1 & 0.01 $\pm$ 0.04 \\ \hline
$Z_2Z_4$ & 0 & 0.00 $\pm$ 0.04 & 0.0 $\pm$ 0.0004 \\ \hline
$Z_2Z_4Z_5$ & 0 & 0.00 $\pm$ 0.01 & 0.0 $\pm$ 0.00006 \\ \hline
$Z_2Z_3$ & 0 & 0.0 $\pm$ 0.1 & 0.0 $\pm$ 0.003 \\ \hline
$Z_2Z_3Z_5$ & 0 & 0.01 $\pm$ 0.02 & 0.0002 $\pm$ 0.0009 \\ \hline
$Z_2Z_3Z_4$ & 0 & -0.002 $\pm$ 0.006 & 0.0 $\pm$ 0.00003 \\ \hline
$Z_2Z_3Z_4Z_5$ & 0 & 0.001 $\pm$ 0.005 & 0.0 $\pm$ 0.00001 \\ \hline
$Z_1$ & 0 & -0.02 $\pm$ 0.04 & 0.0 $\pm$ 0.001 \\ \hline
$Z_1Z_5$ & 0 & 0.000 $\pm$ 0.007 & 0.0 $\pm$ 0.0006 \\ \hline
$Z_1Z_4$ & 0 & 0.000 $\pm$ 0.004 & 0.0 $\pm$ 0.0000006 \\ \hline
$Z_1Z_4Z_5$ & 0 & 0.000 $\pm$ 0.006 & 0.0 $\pm$ 0.00001 \\ \hline
$Z_1Z_3$ & 0 & -0.002 $\pm$ 0.006 & 0.0 $\pm$ 0.00003 \\ \hline
$Z_1Z_3Z_5$ & 0 & -0.001 $\pm$ 0.004 & 0.0 $\pm$ 0.00001 \\ \hline
$Z_1Z_3Z_4$ & 0 & 0.004 $\pm$ 0.005 & 0.00001 $\pm$ 0.00004 \\ \hline
$Z_1Z_3Z_4Z_5$ & 0 & 0.003 $\pm$ 0.004 & 0.00001 $\pm$ 0.00003 \\ \hline
$Z_1Z_2$ & 0 & -0.02 $\pm$ 0.03 & 0.0 $\pm$ 0.001 \\ \hline
$Z_1Z_2Z_5$ & 0 & 0.000 $\pm$ 0.008 & 0.0 $\pm$ 0.0000004 \\ \hline
$Z_1Z_2Z_4$ & 0 & 0.000 $\pm$ 0.005 & 0.0 $\pm$ 0.000004 \\ \hline
$Z_1Z_2Z_4Z_5$ & 0 & 0.000 $\pm$ 0.006 & 0.0 $\pm$ 0.000006 \\ \hline
$Z_1Z_2Z_3$ & 0 & -0.001 $\pm$ 0.007 & 0.0 $\pm$ 0.00002 \\ \hline
$Z_1Z_2Z_3Z_5$ & 0 & 0.000 $\pm$ 0.004 & 0.0 $\pm$ 0.000008 \\ \hline
$Z_1Z_2Z_3Z_4$ & 0 & 0.004 $\pm$ 0.003 & 0.00001 $\pm$ 0.00003 \\ \hline
$Z_1Z_2Z_3Z_4Z_5$ & 0 & 0.004 $\pm$ 0.003 & 0.00001 $\pm$ 0.00003 \\ \hline
\end{tabular}
}
\caption{Expectation value of observables $M$ computed with the exact solution, and with the output solution of  VQLS. $D(M)$ measures the difference between these two results. The linear system considered is $A_{32\times32} = \id + 0.2 X_1Z_2 + 0.2 X_1$, and $\ket{b} = H_1H_3H_4H_5\ket{\vec{0}}$.  \label{fig:5qtable3terms}}
\end{table}

\clearpage
\newpage

\end{document}